\documentclass[a4paper,11pt]{article}
\usepackage{jcappub}

\usepackage{amssymb}
\usepackage{graphicx}

\usepackage{bm}
\usepackage{amsmath}
\usepackage{subcaption}
\usepackage{wrapfig}
\usepackage{sidecap}
\usepackage[separate-uncertainty]{siunitx}
\usepackage{multirow}
\usepackage{orcidlink}

\usepackage{hyperref}
\usepackage[nameinlink,noabbrev]{cleveref}

\def\setsymbol#1#2{\expandafter\def\csname #1\endcsname{#2}}
\def\getsymbol#1{\csname #1\endcsname}

\def\Planck{\textit{Planck}}

\newbox\tablebox    \newdimen\tablewidth
\def\leaderfil{\leaders\hbox to 5pt{\hss.\hss}\hfil}
\def\endPlancktable{\tablewidth=\columnwidth 
    $$\hss\copy\tablebox\hss$$
    \vskip-\lastskip\vskip -2pt}

\def\tablenote#1 #2\par{\begingroup \parindent=0.8em
    \abovedisplayshortskip=0pt\belowdisplayshortskip=0pt
    \noindent
    $$\hss\vbox{\hsize\tablewidth \hangindent=\parindent \hangafter=1 \noindent
    \hbox to \parindent{$^#1$\hss}\strut#2\strut\par}\hss$$
    \endgroup}
\def\doubleline{\vskip 3pt\hrule \vskip 1.5pt \hrule \vskip 5pt}

\def\L2{\ifmmode L_2\else $L_2$\fi}

\def\DeltaT{\ifmmode \Delta T\else $\Delta T$\fi}
\def\deltat{\ifmmode \Delta t\else $\Delta t$\fi}
\def\fknee{\ifmmode f_{\rm knee}\else $f_{\rm knee}$\fi}
\def\Fmax{\ifmmode F_{\rm max}\else $F_{\rm max}$\fi}
\def\solar{\ifmmode{\rm M}_{\mathord\odot}\else${\rm M}_{\mathord\odot}$\fi}
\def\Msolar{\ifmmode{\rm M}_{\mathord\odot}\else${\rm M}_{\mathord\odot}$\fi}
\def\Lsolar{\ifmmode{\rm L}_{\mathord\odot}\else${\rm L}_{\mathord\odot}$\fi}
\def\inv{\ifmmode^{-1}\else$^{-1}$\fi}
\def\mo{\ifmmode^{-1}\else$^{-1}$\fi}
\def\sup#1{\ifmmode ^{\rm #1}\else $^{\rm #1}$\fi}
\def\expo#1{\ifmmode \times 10^{#1}\else $\times 10^{#1}$\fi}
\def\,{\thinspace}
\def\lsim{\mathrel{\raise .4ex\hbox{\rlap{$<$}\lower 1.2ex\hbox{$\sim$}}}}
\def\gsim{\mathrel{\raise .4ex\hbox{\rlap{$>$}\lower 1.2ex\hbox{$\sim$}}}}

\def\simprop{\mathrel{\raise .4ex\hbox{\rlap{$\propto$}\lower 1.2ex\hbox{$\sim$}}}}
\def\deg{\ifmmode^\circ\else$^\circ$\fi}
\def\pdeg{\ifmmode $\setbox0=\hbox{$^{\circ}$}\rlap{\hskip.11\wd0 .}$^{\circ}
          \else \setbox0=\hbox{$^{\circ}$}\rlap{\hskip.11\wd0 .}$^{\circ}$\fi}
\def\arcs{\ifmmode {^{\scriptstyle\prime\prime}}
          \else $^{\scriptstyle\prime\prime}$\fi}
\def\arcm{\ifmmode {^{\scriptstyle\prime}}
          \else $^{\scriptstyle\prime}$\fi}
\newdimen\sa  \newdimen\sb
\def\parcs{\sa=.07em \sb=.03em
     \ifmmode \hbox{\rlap{.}}^{\scriptstyle\prime\kern -\sb\prime}\hbox{\kern -\sa}
     \else \rlap{.}$^{\scriptstyle\prime\kern -\sb\prime}$\kern -\sa\fi}
\def\parcm{\sa=.08em \sb=.03em
     \ifmmode \hbox{\rlap{.}\kern\sa}^{\scriptstyle\prime}\hbox{\kern-\sb}
     \else \rlap{.}\kern\sa$^{\scriptstyle\prime}$\kern-\sb\fi}
\def\ra[#1 #2 #3.#4]{#1\sup{h}#2\sup{m}#3\sup{s}\llap.#4}
\def\dec[#1 #2 #3.#4]{#1\deg#2\arcm#3\arcs\llap.#4}
\def\deco[#1 #2 #3]{#1\deg#2\arcm#3\arcs}
\def\rra[#1 #2]{#1\sup{h}#2\sup{m}}

\def\dots{\relax\ifmmode \ldots\else $\ldots$\fi}
\def\WHzsr{\ifmmode $W\,Hz\mo\,sr\mo$\else W\,Hz\mo\,sr\mo\fi}
\def\mHz{\ifmmode $\,mHz$\else \,mHz\fi}
\def\GHz{\ifmmode $\,GHz$\else \,GHz\fi}
\def\mKs{\ifmmode $\,mK\,s$^{1/2}\else \,mK\,s$^{1/2}$\fi}
\def\muKs{\ifmmode \,\mu$K\,s$^{1/2}\else \,$\mu$K\,s$^{1/2}$\fi}
\def\muKRJs{\ifmmode \,\mu$K$_{\rm RJ}$\,s$^{1/2}\else \,$\mu$K$_{\rm RJ}$\,s$^{1/2}$\fi}
\def\muKHz{\ifmmode \,\mu$K\,Hz$^{-1/2}\else \,$\mu$K\,Hz$^{-1/2}$\fi}
\def\MJysr{\ifmmode \,$MJy\,sr\mo$\else \,MJy\,sr\mo\fi}
\def\MJysrmK{\ifmmode \,$MJy\,sr\mo$\,mK$_{\rm CMB}\mo\else \,MJy\,sr\mo\,mK$_{\rm CMB}\mo$\fi}
\def\microns{\ifmmode \,\mu$m$\else \,$\mu$m\fi}

\def\muK{\ifmmode \,\mu$K$\else \,$\mu$\hbox{K}\fi}
\def\microK{\ifmmode \,\mu$K$\else \,$\mu$\hbox{K}\fi}
\def\muW{\ifmmode \,\mu$W$\else \,$\mu$\hbox{W}\fi}
\def\kms{\ifmmode $\,km\,s$^{-1}\else \,km\,s$^{-1}$\fi}
\def\kmsMpc{\ifmmode $\,\kms\,Mpc\mo$\else \,\kms\,Mpc\mo\fi}
\providecommand{\sorthelp}[1]{}

\def\SEVEM{{\tt SEVEM}}
\def\sevem{{\tt SEVEM}}
\def\commander{\texttt{Commander}}
\def\Commander{\texttt{Commander}}

\newcommand\planck{{\it Planck}}
\newcommand\nside{$N_\mathrm{side}$}

\crefname{equation}{eq.}{eqs.}
\Crefname{equation}{Equation}{Equations}

\DeclareSIUnit{\KRJ}{K_\mathrm{RJ}}

\title{Planck PR4 (NPIPE) map-space cosmic birefringence}

\author[a,b]{Raelyn M. Sullivan\,\orcidlink{0000-0003-3819-7526},}
\author[a]{Arefe Abghari\,\orcidlink{0009-0003-2250-3880},}
\author[c]{Patricia Diego-Palazuelos\,\orcidlink{0000-0002-5129-2379},}
\author[a]{Lukas\,T.\,Hergt\,\orcidlink{0000-0003-2730-7419},}
\author[a]{Douglas Scott\,\orcidlink{0000-0002-6878-9840}}

\affiliation[a]{
    Department of Physics and Astronomy,
    University of British Columbia,
    6224 Agricultural Road,
    Vancouver, BC V6T 1Z1, Canada
}
\affiliation[b]{Institute for Theoretical Astrophysics, University of Oslo, Blindern, Oslo, Norway}
\affiliation[c]{Max-Planck-Institut für Astrophysik,
Karl-Schwarzschild-Str. 1,
85748 Garching, Germany}

\emailAdd{rsullivan@phas.ubc.ca}

\date{\today}

\abstract{
 Cosmic birefringence is an effect where the plane of polarisation of the cosmic microwave background (CMB) is rotated by an angle $\beta$ through coupling to a hypothetical parity-violating field.
 We analyse the \planck\ Public Release 4 (PR4 or NPIPE) data using a map-space analysis method and find $\beta=\ang{0.46}\pm\ang{0.04}(\mathrm{stat.})\pm\ang{0.28}(\mathrm{syst.})$ for \SEVEM\ CMB maps and $\beta=\ang{0.48}\pm\ang{0.04}(\mathrm{stat.})\pm\ang{0.28}(\mathrm{syst.})$ for \Commander\ CMB maps. 
These values are slightly higher than previously published results, which may be explained by the fact that we have not attempted to remove any potential bias from miscalibration of the \planck\ polarimeters.
The uncertainty in this miscalibration dominates the systematic uncertainty, which also means that our results are consistent with no parity violation. An advantage of the map-space analysis is that it is easy to investigate any variations on the sky, for example caused by foreground contamination. Our results for isotropic birefringence are fairly robust against different spatial data cuts, but there may be hints of a foreground systematic (north versus south hemispheres) or uncontrolled miscalibration effect ($T$ peaks versus $E$ peaks) that should be followed up in future studies. We additionally find no
evidence of a cosmic birefringence dipole (anisotropic birefringence).}

\begin{document}
\maketitle
\flushbottom

\section{Introduction}
\label{sec:intro}

Cosmic birefringence is a generic result of adding a parity-violating term to the electromagnetic Lagrangian \cite{1990CarrollBiref,2022Komatsu}, as might occur in axionic dark matter \cite{2017Liu_axionDM,2023Zhou_axionDM,2022Obata_axionDM,2021Nakagawa_axionDM} or dark energy \cite{2021Choi_axionLDE,2018Poulin_axionLDE,2021Fujita_axionLDE,2023Kamionkowski_axionEDE, 2023EDE,2023EDE2,2024EDE}. A consequence of this modification to the electromagnetic Lagrangian is a slow, persistent rotation of the polarisation angle for photons travelling through the Universe. For a photon emitted at the last scattering surface, this results in a characteristic rotation angle that we denote $\beta$.

\begin{SCfigure}[2][tbp!]
    \centering
    \includegraphics[width=0.4\linewidth]{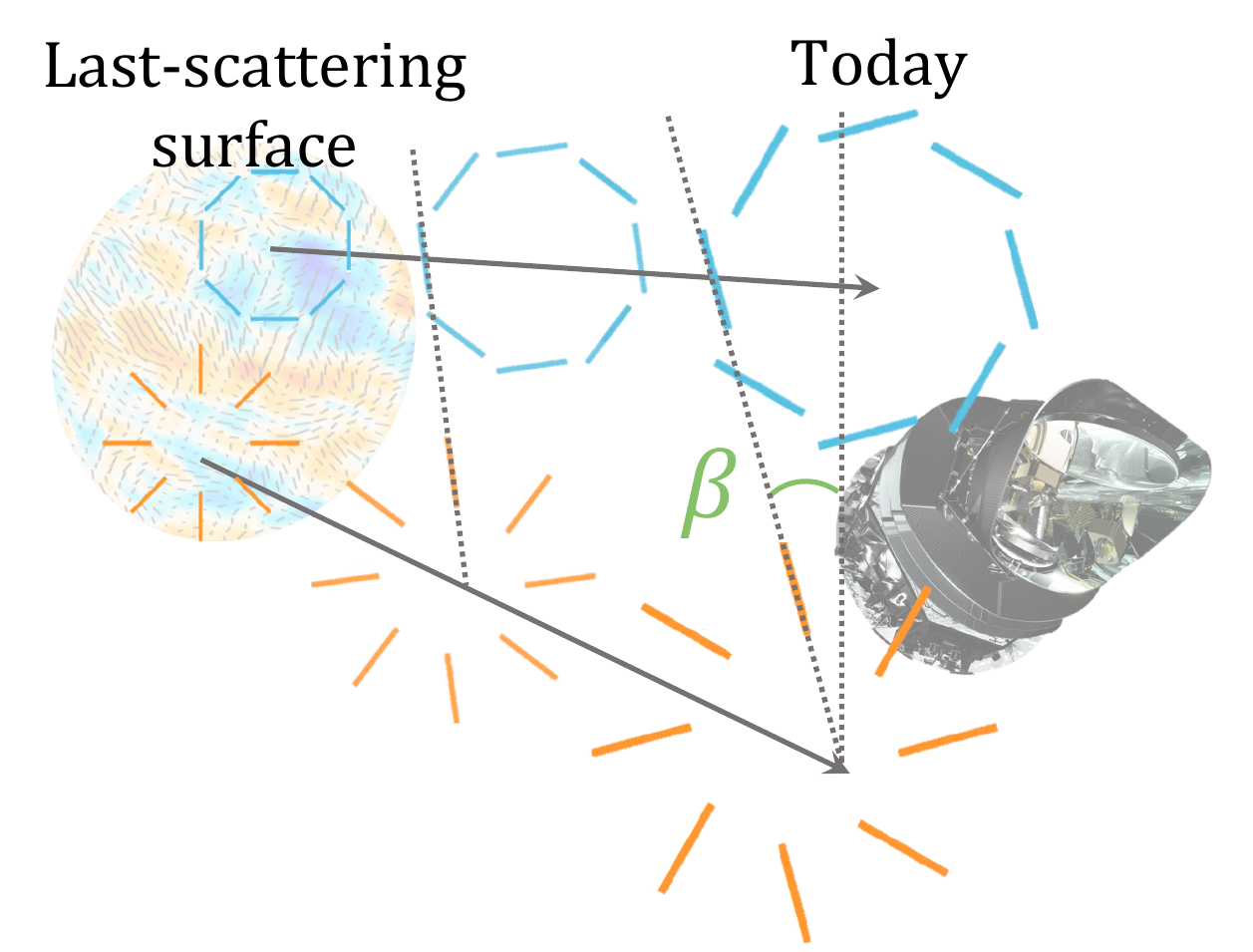}
    \caption{Schematic of cosmic birefringence. In many models the effect accumulates over the distance travelled, so the largest signal would be observed in the oldest (or furthest travelled) photons. This makes the CMB the ideal signal to search for cosmic birefringence. The effect rotates the plane of polarisation of the photons as they travel, so that there will be a small amount of $EB$ cross-correlation in the final rotated polarisation vectors, which would be expected to be zero in the case of no cosmic birefringence. The signal here is exaggerated for clarity, with $\beta=\ang{15}$.}
    \label{fig:birefrotschematic}
\end{SCfigure}

This rotation mixes the curl-free $E$ modes with the gradient-free $B$ modes from the cosmic microwave background (CMB) and produces $EB$ and $TB$ correlations ($T$ for Stokes $I$, intensity or temperature), which are absent without such a parity-violating effect. 
The simplest models producing cosmic birefringence find that this rotation accumulates with the propagation distance of the photons~\cite{2023Galaverni, 2024Naokawa}, making the photons from the last-scattering surface (the CMB) the ideal source for studying this phenomenon \cite{2022Komatsu},
as shown schematically in \cref{fig:birefrotschematic}.

This effect has already been heavily investigated, most analyses being consistent with no birefringence \cite{2018CBRadio,2024CBRadio,SPTBiref,sptbiref2,bicepbiref,bicepbiref2,polarbearBiref}; however, tantalizing hints of a signal were made using the \Planck\ Public Release 2 (PR2) data \cite{2017Contreras,planck2014-a23} of $\beta=\ang{0.31}\pm\ang{0.05} (\mathrm{stat.})\pm\ang{0.28}\mathrm{(syst.)}$ from a harmonic analysis and $\beta=\ang{0.35}\pm\ang{0.05}(\mathrm{stat.})\pm\ang{0.28}\mathrm{(syst.)}$ from a map analysis.  The results were below the uncertainty level for the miscalibration angle of the \Planck\ polarimeters~\cite{rosset2010,planck2014-a09} and thus compatible with no birefringence. The effect of cosmic birefringence on the CMB power spectrum is degenerate (or nearly degenerate, depending on the model tested \cite{2023Sherwin,2022Nakatsuka,2023Galaverni,2024Greco}), with a rotational offset in the telescope polarisation detectors. 
Since the inflight calibration of detectors requires precise prior knowledge on the emission of reference astrophysical sources (e.g., the Crab Nebula~\cite{2018Ritacco, 2020Aumont}) and the instrument optics, this dramatically reduces the detection potential for satellite experiments. Ground-based experiments can instead rely on artificial calibration sources to break the degeneracy between instrumental and cosmological rotations~\cite{2022Cornelison, 2024SO2, 2023SO1, 2017Polocalc, 2021SO4, 2024SO3}.

A follow-up study on the \Planck\ Public Release 3 and 4 data \cite{2020Yuto-biref,PR4biref} effectively used the foreground $EB$ signal to constrain the effect of the miscalibration, finding an angle consistent with $\ang{0.30}\pm\ang{0.11}(\mathrm{stat.+syst.})$; however, no cosmological significance was attributed to the signal since it varied with the use of different masks, to as low as  $\ang{0.10}\pm\ang{0.21}$ for $f_\mathrm{sky}=0.75$ to as high as $\ang{0.36}\pm\ang{0.11}$ for $f_\mathrm{sky}=0.93$. 
This cast doubt on the accuracy of the available foreground $EB$ models and on the methodology's ability to disentangle cosmic birefringence from instrumental systematics.
Although these values are consistent within the uncertainty, they motivate further study of the origin of $\beta$, whether cosmological or instrumental. 
Since then there has been a flurry of activity trying to analyse the data sets more carefully, to determine whether this signal can be attributed to telescope systematic effects, foregrounds, modelling, or new physics and whether we can squeeze any more significance from the data we currently have \cite{PlanckWmap-biref,EskiltLFIHFI-biref,2022Abghari,PR4biref-diegoPat,Cosmoglobe-biref,2023Freq,2022Fujita,2023EDE,2024EDE,2024magdust2}.

Several studies have also investigated anisotropic birefringence~\cite{2009Kamionkowski}. In this case, $\beta$ varies across the sky, and should not be affected by the miscalibration angle (expected to be isotropic on the sky) and have found null detections \cite{2017Contreras, 2020Gruppuso, 2022Bortolami, 2024Zagatti, 2025Namikawa}. 
As well as the all-sky signals that we focus on here, there have also been many studies looking at ground-based experiments, none finding any significant detection of isotropic or anisotropic cosmic birefringence \cite{polarbearBiref,bicepbiref,bicepbiref2,actBiref,SPTBiref,sptbiref2}.

In numerous CMB studies, it has been common practice to test results found using CMB power spectra using alternative map-space methods. Since each analysis is affected by different systematic effects, this can be a fruitful way to confirm or refute results from prior analyses. The map-space analysis also easily allows us to check masking effects, on areas with greater foreground contamination, for example.  In the study of cosmic birefringence, a map-space analysis also allows us to easily search for an anisotropic birefringence signal, since the spatially varying component is naturally found through the variation of the measured signal over the sky.

The isotropic birefringence signal alters the CMB power spectra such that a birefringence angle of $\beta$ rotates the CMB angular power spectra (see e.g.~\cite{Lue:1998mq,Feng:2004mq,Liu:2006uh}) according to
\begin{subequations}
    \label{eq:cls_biref}
    \begin{align}
        C_\ell^{TT,\mathrm{obs}}&=C_\ell^{TT}\, , \label{eq:cls_biref1} \\ 
        C_\ell^{EE,\mathrm{obs}}&=C_\ell^{EE}\cos^2(2\beta)+C_\ell^{BB}\sin^2(2\beta)\, ,\label{eq:cls_biref2}\\ 
        C_\ell^{BB,\mathrm{obs}}&=C_\ell^{EE}\sin^2(2\beta)+C_\ell^{BB}\cos^2(2\beta)\, ,\label{eq:cls_biref3}\\ 
        C_\ell^{TE,\mathrm{obs}}&=C_\ell^{TE}\cos(2\beta)\, ,\label{eq:cls_biref4}\\ 
        C_\ell^{TB,\mathrm{obs}}&=C_\ell^{TE}\sin(2\beta)\, ,\label{eq:cls_biref5}\\ 
        C_\ell^{EB,\mathrm{obs}}&=\frac{1}{2}\left(C_\ell^{EE}-C_\ell^{BB}\right)\sin(4\beta)\, ,
        \label{eq:cls_biref6}
    \end{align}
\end{subequations}
\noindent where $C_\ell^{XX,\mathrm{obs}}$ are the observed (rotated) power spectra for $X\in{T,E,B}$.
Since the expected cosmological $C_\ell^{EB,\mathrm{obs}}$ and $C_\ell^{TB,\mathrm{obs}}$ signals are zero in the absence of cosmic birefringence, these are the signatures of greatest interest for this study. 

In \cref{sec:data} we outline the data and simulations used for the analysis, described in \cref{sec:method}. We will explore the results of the map-space analysis method in \cref{sec:res} and touch on measures of the anisotropic birefringence there. We will conclude in \cref{sec:birefconc}.

\section{Data and simulations}
\label{sec:data}
For this analysis, we use the data collected and analysed by the Planck collaboration\footnote{Based on observations obtained with \Planck\ (\url{http://www.esa.int/Planck}), an ESA science mission with instruments and contributions directly funded by ESA Member States, NASA, and Canada.} and the CMB map analysis software \texttt{HEALPix}~\cite{gorski2005}.\footnote{\url{http://healpix.sf.net}}  
\Planck\ was a space-based CMB experiment observing the microwave sky from 2009 to 2013, in nine frequency bands ranging from 30 to 857$\,$GHz in temperature (Stokes $I$, denoted by $T$) and seven frequency bands from 30 to 353$\,$GHz in polarisation (Stokes $Q$ and $U$). This gave it unprecedented precision for foreground removal and characterization, and the best all-sky observations to date. The Planck Collaboration produced four public data releases and hundreds of publications \cite{planck2013-p01,planck2014-a01,planck2016-l01,planck2020-LVII}. 

\subsection{Data}
\begin{figure*}[tbp!]
     \centering
     \begin{subfigure}[b]{0.49\textwidth}
         \centering
         \includegraphics[width=\textwidth]{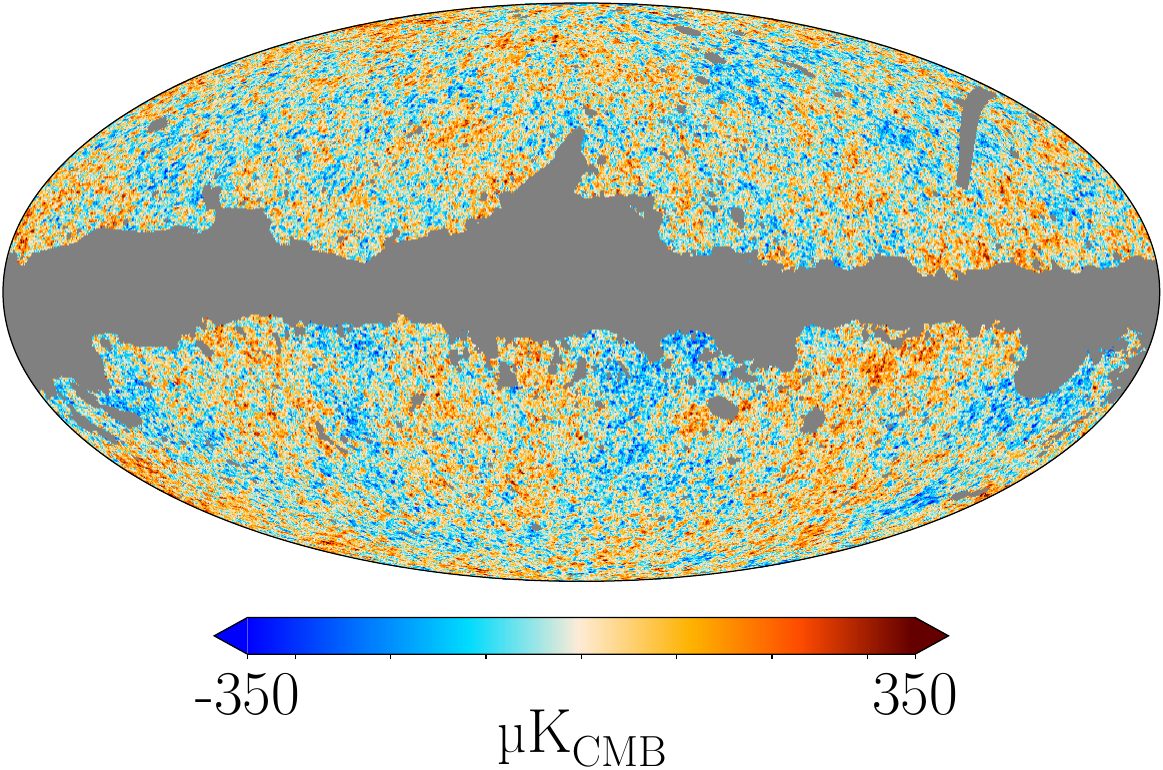}
         \caption{Intensity map}
         \label{fig:dataT}
     \end{subfigure}
     \begin{subfigure}[b]{0.49\textwidth}
         \centering
         \includegraphics[width=\hsize]{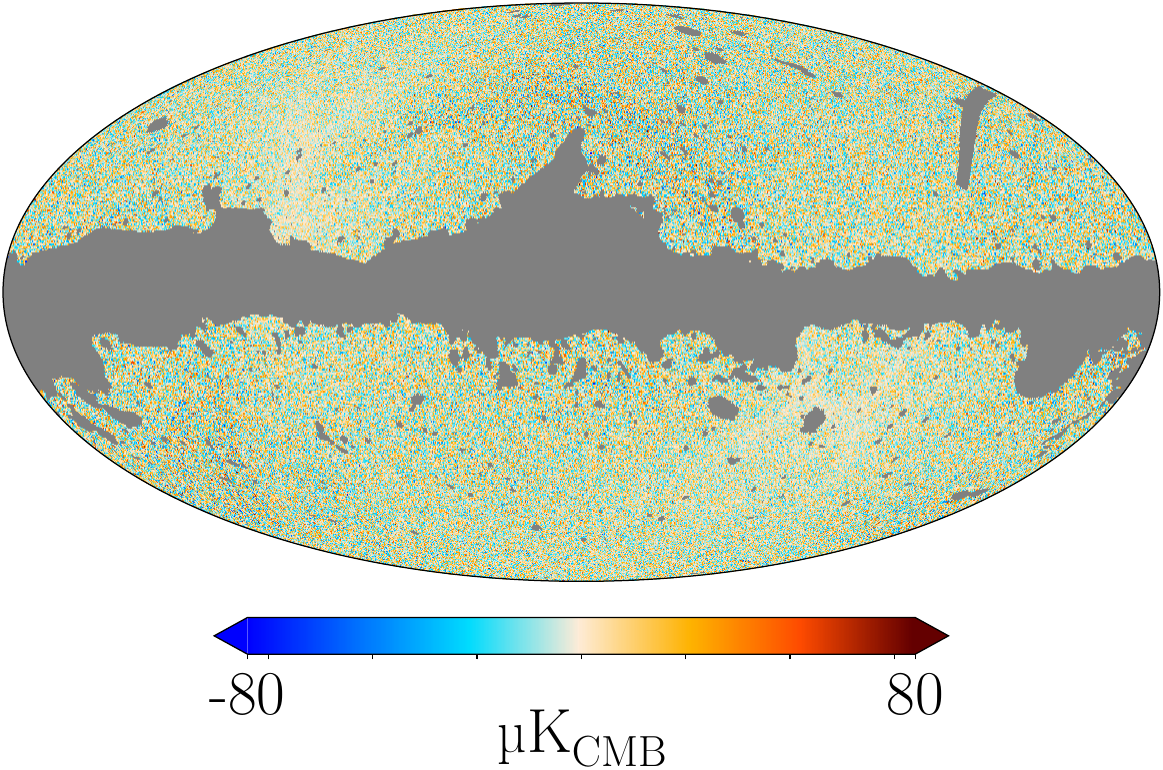}
         \caption{$E$-mode map}
         \label{fig:dataE}
     \end{subfigure}
     \begin{subfigure}[b]{0.49\textwidth}
         \centering
         \includegraphics[width=\hsize]{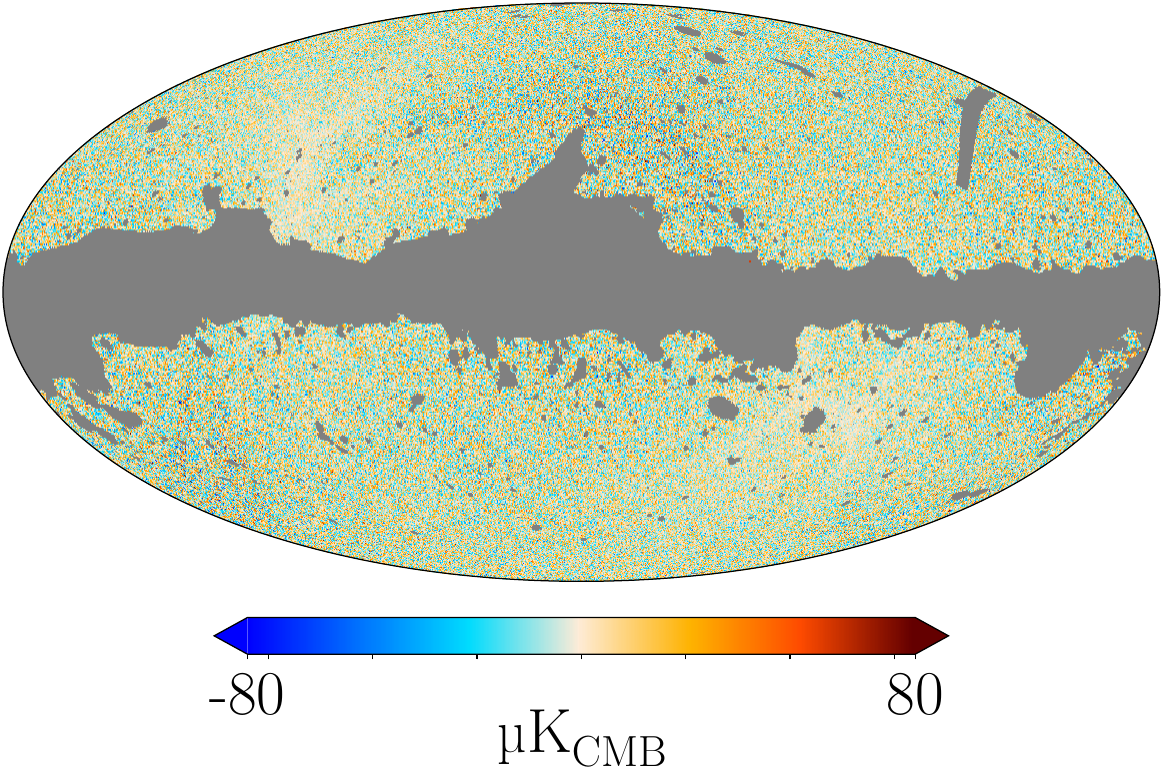}
         \caption{Stokes $Q$}
         \label{fig:dataQ}
     \end{subfigure}
     \begin{subfigure}[b]{0.49\textwidth}
         \centering
         \includegraphics[width=\hsize]{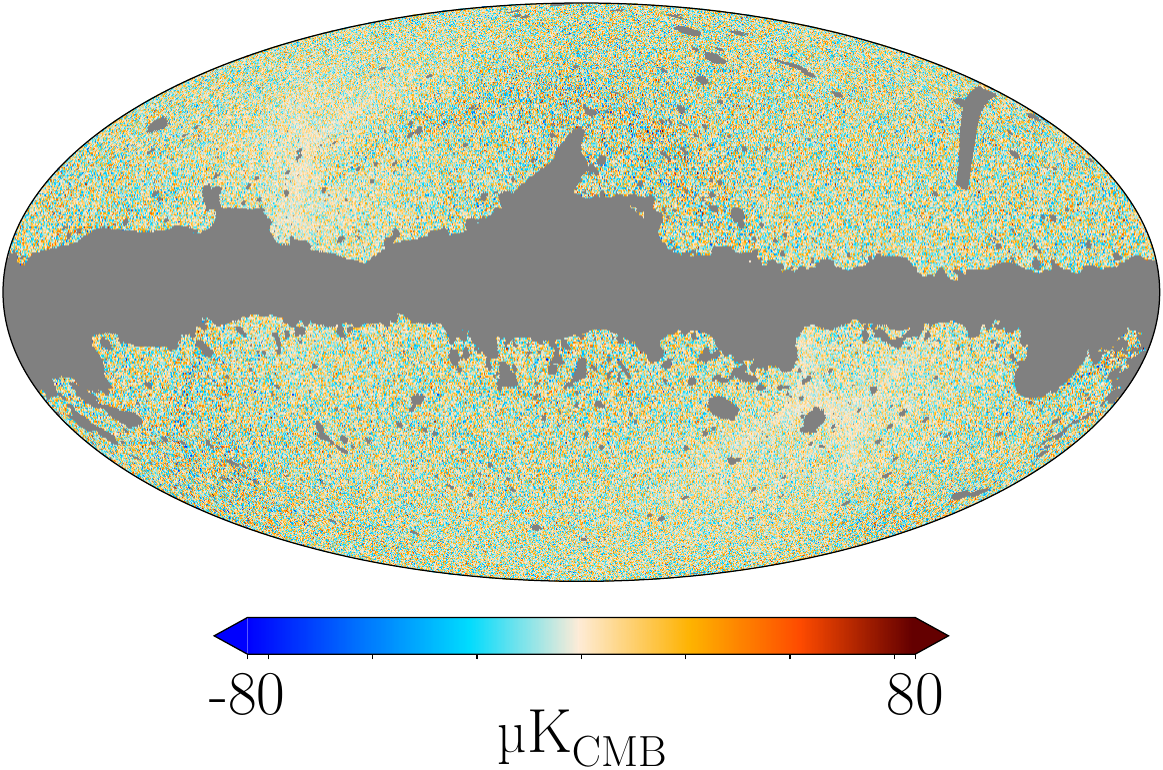}
         \caption{Stokes $U$}
         \label{fig:dataU}
     \end{subfigure}
        \caption{\Planck\ PR4 \SEVEM\ component-separated temperature map, $E$-mode map, and $Q$ and $U$ polarisation maps, masked with the union of the common intensity and polarisation mask ($f_{\mathrm{sky}}=0.75$). Panels (a) and (b) are used to find the temperature and $E$ peaks for the analysis, and (c) and (d) are used to determine $Q_r$ and $U_r$.}
        \label{fig:birefdata}
\end{figure*}

We use the \Planck\ Public Release 4~\cite{planck2020-LVII}, also known as NPIPE, \SEVEM\footnote{Spectral Estimation Via Expectation Maximisation (\SEVEM) \cite{leach2008,2012sevem-pol}.} and \commander\footnote{Commander is an Optimal Monte-carlo Markov chAiN Driven EstimatoR (\commander, \url{https://github.com/Cosmoglobe/Commander}) \cite{2004Jewell, 2004Wandelt, 2004Eriksen, 2008Eriksen}.}
component-separated CMB maps for this analysis. We consider the 300 end-to-end \SEVEM\ simulations and compare them with the results from \SEVEM\ for the real data (see \cref{fig:birefdata}) and the results from \commander. These end-to-end simulations include all known instrumental systematic effects, as well as CMB and foregrounds,\footnote{\sevem\ simulations are free from residual foregrounds left by the component-separation method but do include foreground systematics.} and are the most realistic simulation sets produced by the Planck collaboration \cite{planck2016-l03,planck2020-LVII}. 
The initial resolution of the maps is \nside$=2048$, with a Gaussian beam with a full-width-at-half-maximum (FWHM) of \ang{;5;}. The simulations do not include a birefringence offset, but will have effects from component separation and the instrument that are expected to emulate the true data~\cite{PR4biref-diegoPat}. We also generate 100 simulations with an input birefringence rotation angle of \ang{0.3} and the same noise spectrum as the NPIPE maps, but no foreground systematic effects. Pure white-noise simulations, and cosmic-variance limited simulations are also considered and results for these may be found in \cref{app:CVL_biref}.

\subsection{Masks}
\begin{figure*}[tbp!]
         \centering
     \begin{subfigure}[b]{0.49\textwidth}
         \centering
         \includegraphics[width=\textwidth]{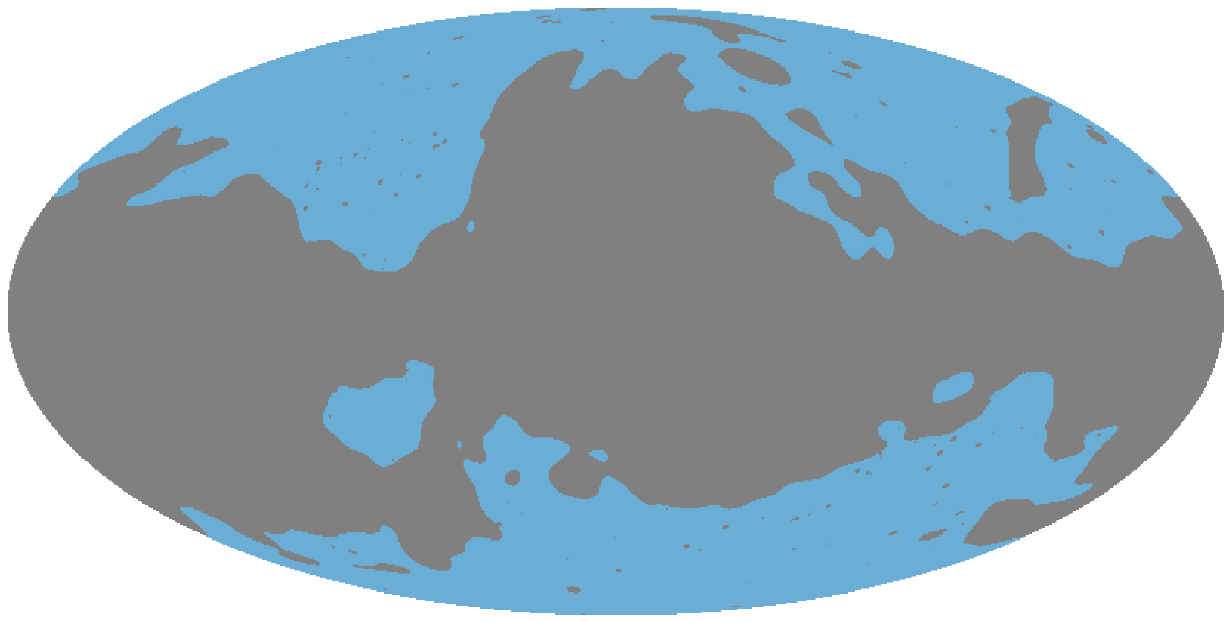}
         \caption{Dust mask, $f_\mathrm{sky}=38\%$.}
         \label{fig:birefdust}
     \end{subfigure}
     \begin{subfigure}[b]{0.49\textwidth}
         \centering
         \includegraphics[width=\hsize]{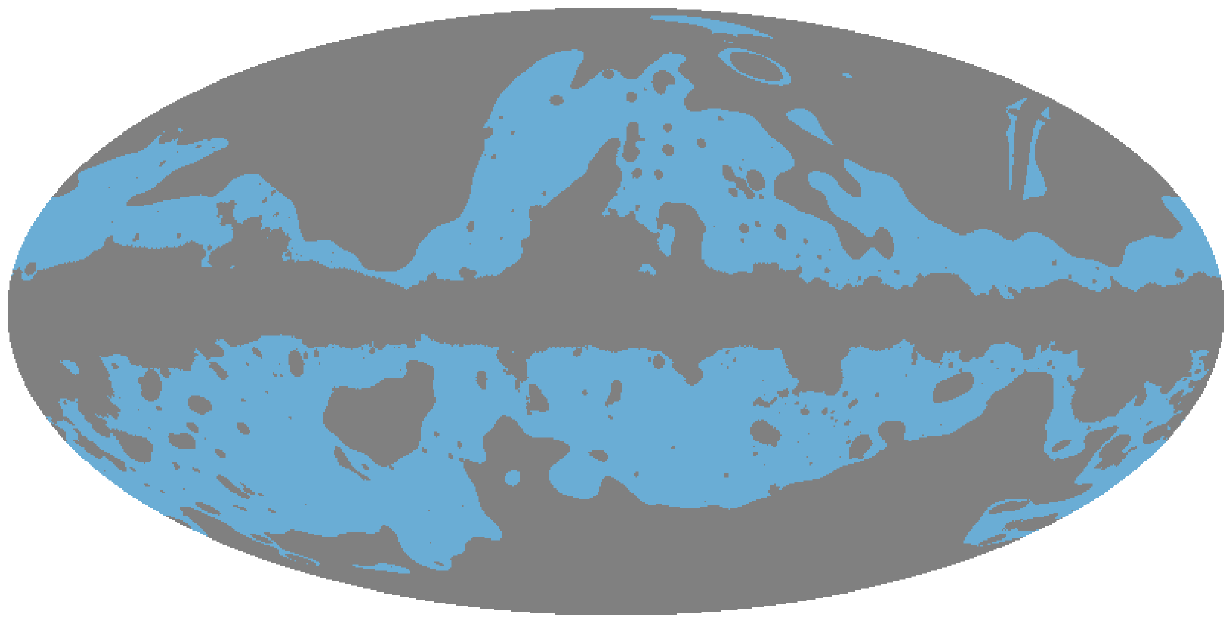}
         \caption{Inverse dust mask, $f_\mathrm{sky}=36\%$. }
         \label{fig:birefidust}
     \end{subfigure}
          \begin{subfigure}[b]{0.49\textwidth}
         \centering
        \includegraphics[width=\textwidth]{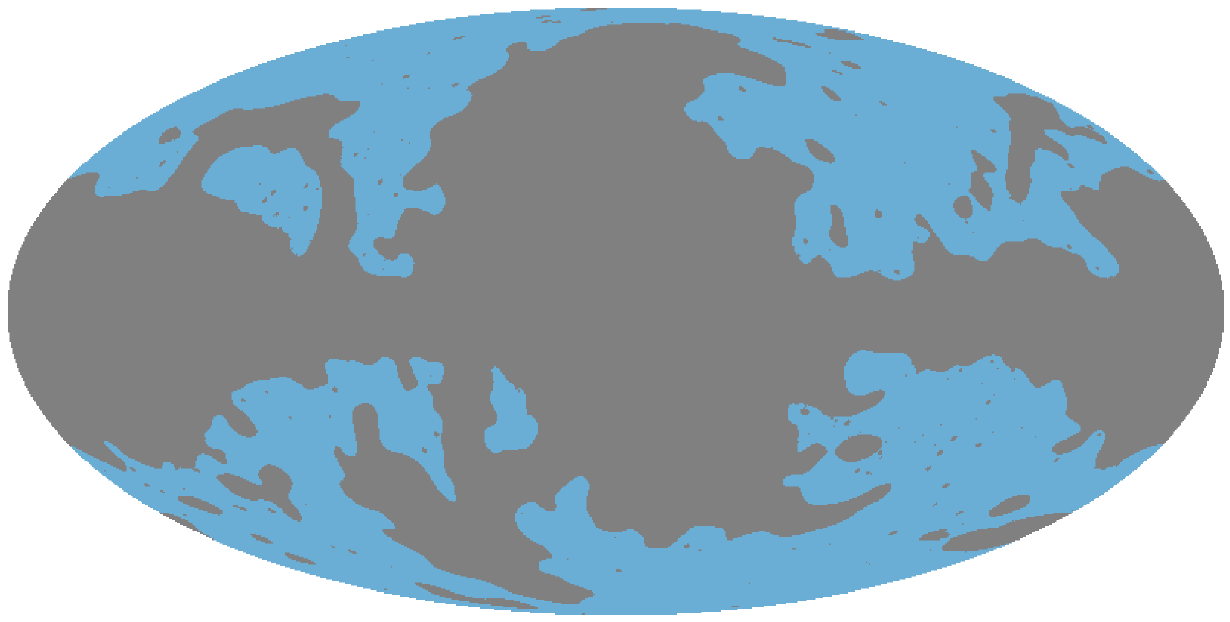}
         \caption{Synchrotron mask, $f_\mathrm{sky}=38\%$.}
         \label{fig:birefsync}
     \end{subfigure}
     \begin{subfigure}[b]{0.49\textwidth}
         \centering
         \includegraphics[width=\hsize]{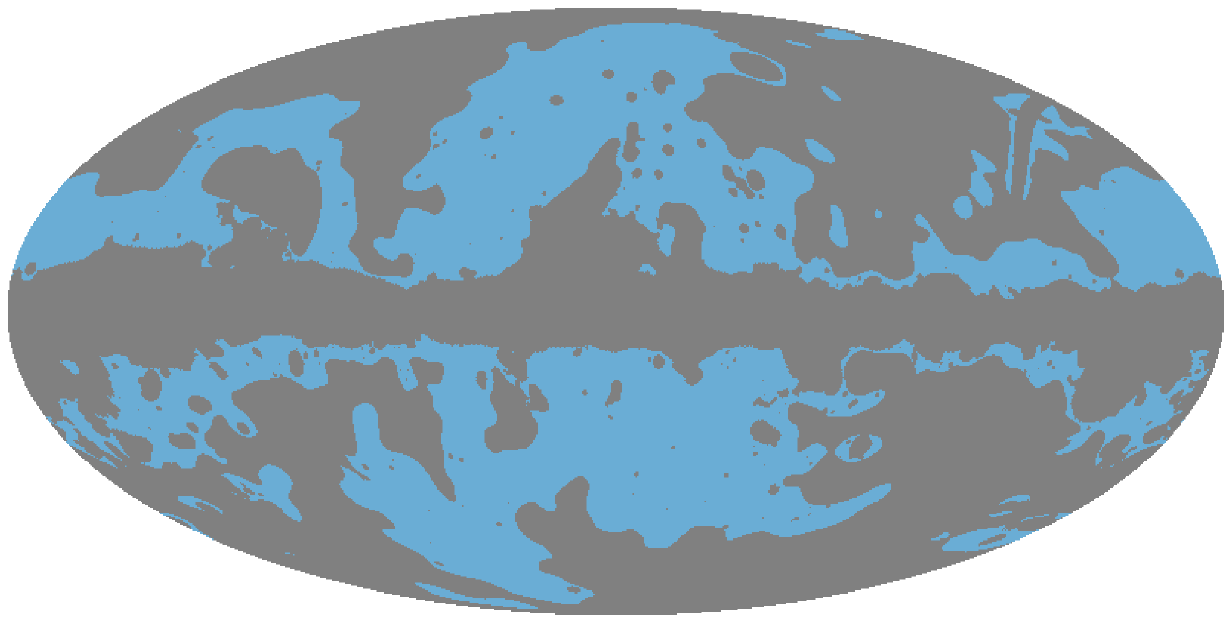}
         \caption{Inverse synchrotron mask, $f_\mathrm{sky}=36\%$. }
         \label{fig:birefisync}
     \end{subfigure}
     \begin{subfigure}[b]{0.49\textwidth}
         \centering
         \includegraphics[width=\hsize]{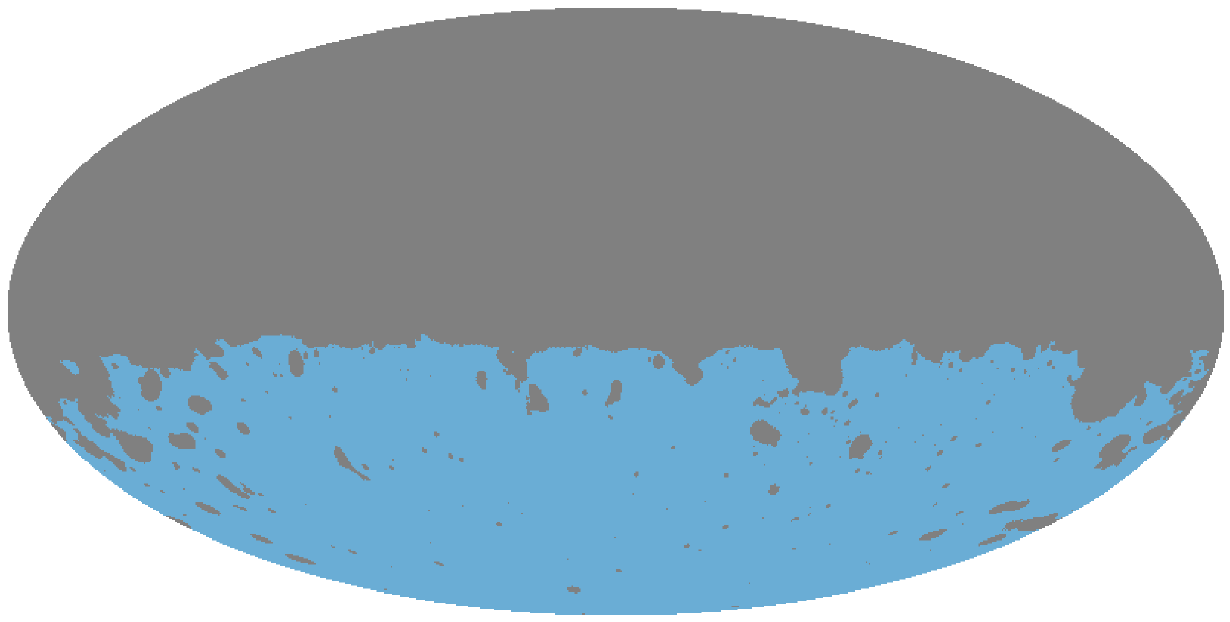}
         \caption{North mask, $f_\mathrm{sky}=37\%$.}
         \label{fig:birefnorth}
     \end{subfigure}
     \begin{subfigure}[b]{0.49\textwidth}
         \centering
         \includegraphics[width=\hsize]{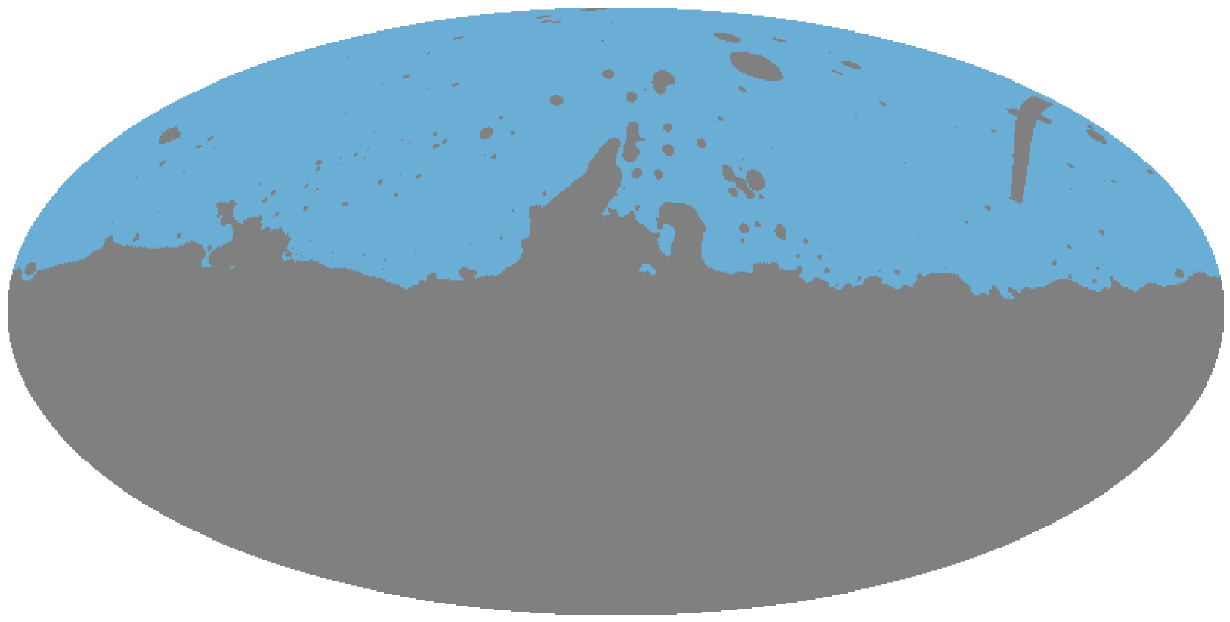}
         \caption{South mask, $f_\mathrm{sky}=38\%$.}
         \label{fig:birefsouth}
     \end{subfigure}
        \caption{Extended masks used for the analysis, with blue showing the region of sky used in each case. The dust and synchrotron masks are set to remove areas with greater dust and synchrotron foregrounds. The inverse dust and synchrotron masks are areas with the highest synchrotron and dust, but with the Galactic plane still removed. Each mask is a union with the common mask, which is applied to each step of the analysis. }
        \label{fig:birefmasks}
\end{figure*}

For most of the analysis we use a union of the common intensity mask and the common polarisation mask (since our analysis concerns the $TB$ and $EB$ correlations), which masks 25\% of the sky,  shown in grey in \cref{fig:birefdata}. 
Since we are interested in whether or not the birefringence angle varies across the sky (in particular concerning the foregrounds), we test several masks in addition to the common mask. We test first a simple northern and southern hemisphere mask (called `North' when the north part is removed, and `South' when the south part is removed, see \cref{fig:birefnorth,fig:birefsouth}), and two more foreground-specific masks, defined from the dust and synchrotron foreground polarisation maps. We first generate a total polarisation $P$ map of synchrotron emission from the $Q$ and $U$ maximum likelihood solutions from the \commander\ component-separation results from PR3. We then choose a threshold of \SI{6.5}{\micro\KRJ}, setting everything above this to zero and below this to one, and smoothed with a \ang{5} FWHM Gaussian beam.\footnote{Rayleigh-Jeans temperature units, also known as brightness temperature, $\mathrm{K}_\mathrm{RJ}$ are conventionally used for the dust and synchrotron foregrounds for numerical stability.} Choosing a threshold of 0.55 on this smoothed map, we set pixels above this value to one and below to zero. The inverse of this map makes up the inverse-synchrotron mask. We then apply the common mask to both the synchrotron and inverse-synchrotron masks (see \cref{fig:birefsync,fig:birefisync}). Dust is treated similarly, this time with an initial threshold of \SI{3.5}{\micro\KRJ}, the same smoothing by \ang{5} and the same final threshold of 0.55 distinguishing the dust from the inverse-dust map (see \cref{fig:birefdust,fig:birefidust}).
Each of these extended masks (North and South, dust and inverse dust, synchrotron and inverse synchrotron) also has the original common mask applied in union with the new mask. These masks can be seen in \cref{fig:birefmasks}. This gives us $f_\mathrm{sky}$ of 75\% for the common mask, 37\% for the North mask, 38\% for the South mask, 38\% for the dust and synchrotron masks, and 36\% for the inverse dust and synchrotron masks.

\section{Map-space analysis method}
\label{sec:method}

\begin{SCfigure}[2.2][tbp!]
    \centering
    \includegraphics[width=0.35\linewidth]{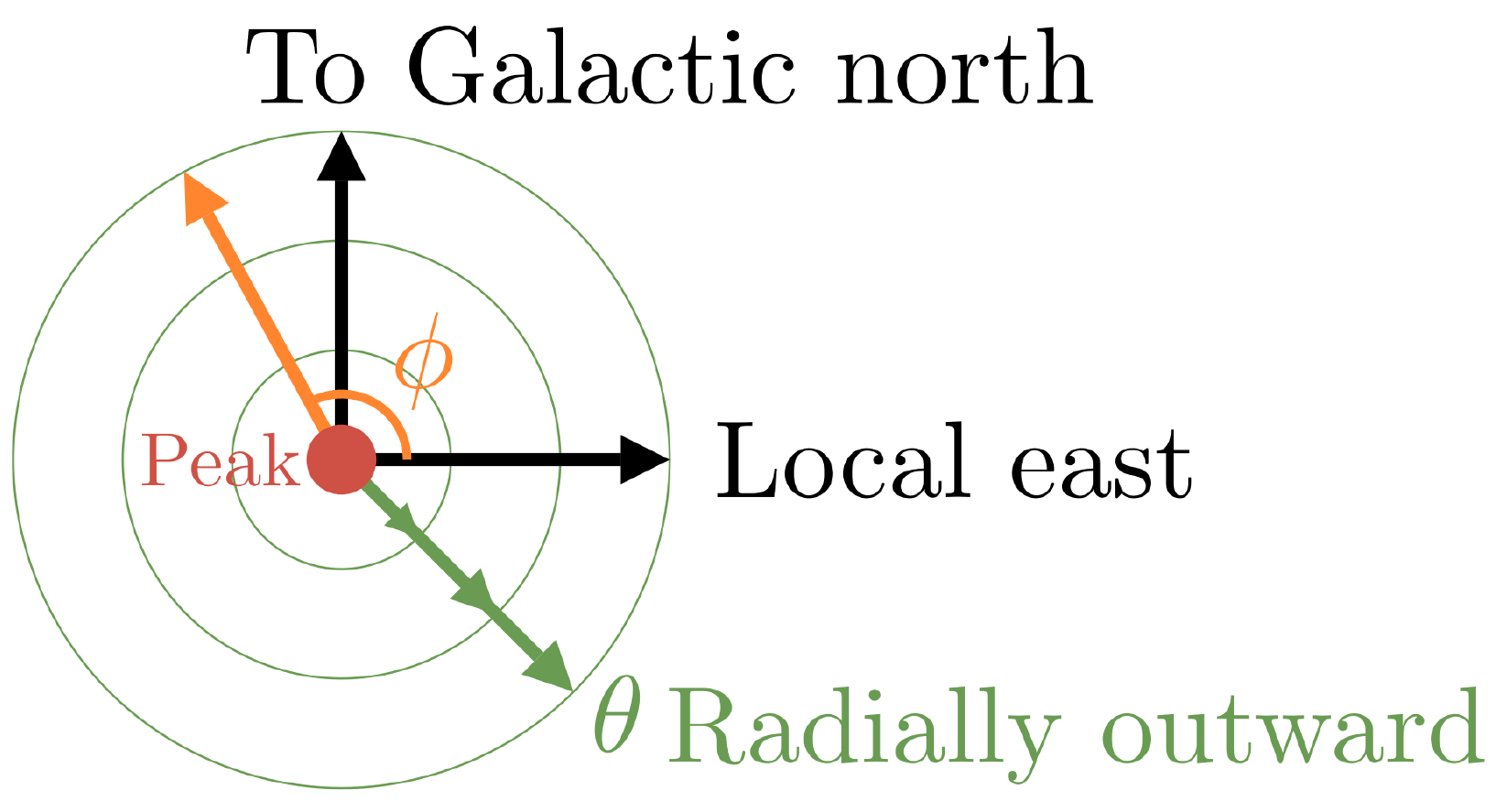}
    \caption{Coordinate system used to transform from the Stokes $Q$ and $U$ parameters to $Q_r$ and $U_r$ (\cref{equ:QrUrExt}). Here $\theta$ (in green) is the angular distance radially outwards from a peak (shown as a red point here), while in orange the angle $\phi$ is defined from a local east direction, chosen such that north points towards the Galactic north along the sphere. }
    \label{fig:biref_coordsys}
\end{SCfigure}

\begin{SCfigure}[2][tbp!]
    \centering
    \includegraphics[width=0.3\linewidth]{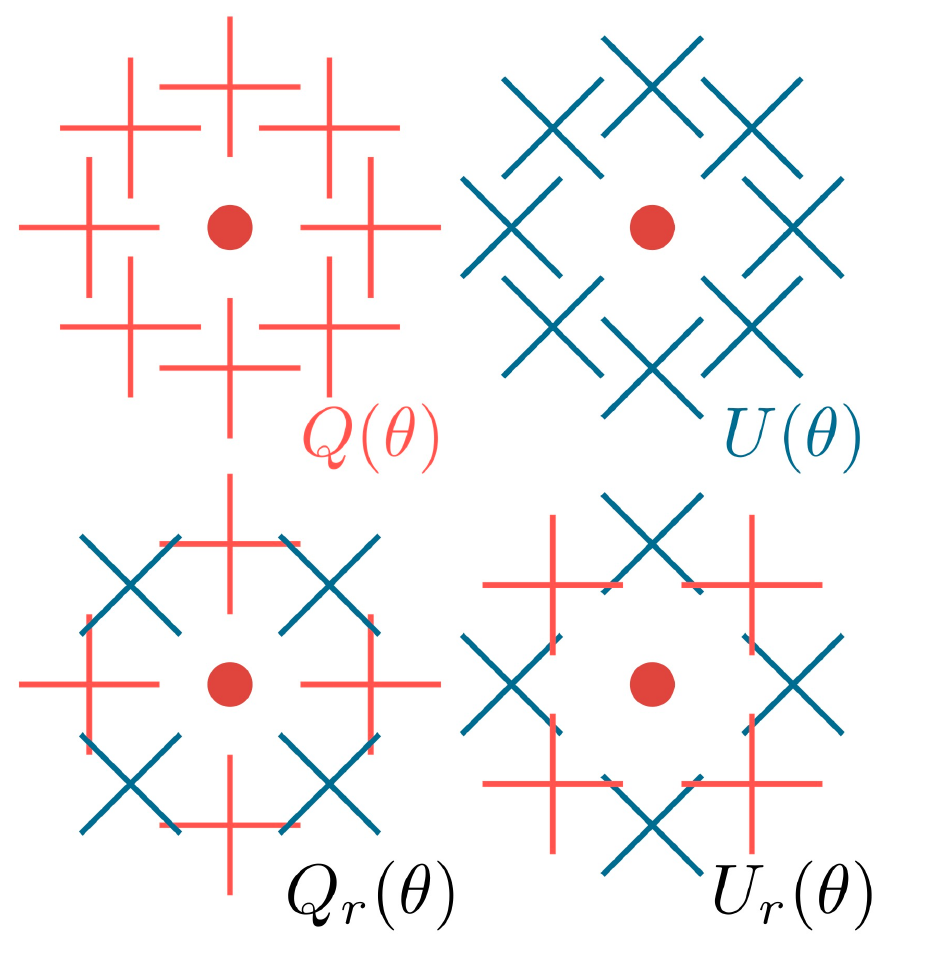}
    \caption{Transforming the Stokes $Q$ and $U$ parameters to $Q_r$ and $U_r$ mixes the modes in such a way as to obtain a local $E$- and $B$-mode measurement around a peak (represented here with the red dot). This mixing pattern is due to the $\cos(2\phi)$ and $\sin(2\phi)$ terms in \cref{equ:QrUrExt}. $Q_r$ shows a clear $E$ pattern (curl-free) whereas $U_r$ shows a clear $B$ pattern. This allows us to search for $E$- and $B$-specific signals in map space, similar to how we might analyse data for $E$- or $B$-specific signals in harmonic space.}
    \label{fig:birefQuQrUr}
\end{SCfigure}

In this section, we describe the steps to measure the cosmic birefringence signal using a map-space peak-stacking method and building upon methods outlined in Refs.~\cite{komatsu2010,planck2014-a23, 2017Contreras}.
We evaluate the $TB$ and $EB$ correlations at extrema using the modified Stokes parameters~\cite{1997Kamiokowski}, transforming the observed $Q$ and $U$ to 
\begin{align}
    Q_r(\theta)&=-Q(\theta)\cos(2\phi)-U(\theta)\sin(2\phi)\, ,\\
    U_r(\theta)&=\phantom{-}Q(\theta)\sin(2\phi)-U(\theta)\cos(2\phi)\, ,   
    \label{equ:QrUrExt}
\end{align}
where $\theta$ is the angular distance radially from the extremum and $\phi$ is defined as the angle from a local east (where north points towards the Galactic north poles), as shown in \cref{fig:biref_coordsys,fig:birefQuQrUr}.
In this way, $Q_r(\theta)$ measures the local $E$ modes at that extremum, and $U_r(\theta)$ measures the local $B$ modes. 
The extrema are found in the $T$ and $E$ maps, to obtain the equivalent of the $TB$ and $EB$ correlations when considering $U_r(\theta)$ \cite{1997Kamiokowski}. Equivalently, the $TE$ and $EE$ correlations may be found by considering $Q_r(\theta)$ at the $T$ and $E$ extrema, but since these have very little constraining power on cosmic birefringence \cite{2022Abghari} they will not be considered further here.

We can determine the expected profile around a peak in a Gaussian random field by considering peak theory, following the work of Ref.~\cite{1986Bardeen} for 3-dimensional fields and Refs.~\cite{1987Bond,2008Desjacques} in 2-dimensions, as well as derivations in Ref.~\cite{komatsu2010}. From these analyses we obtain the number density contrast, $d_\mathrm{pk}(\hat{n})$, at position $\hat{n}$:
\begin{subequations}
\label{eq:bias_params_b}
    \begin{align}
    d_\mathrm{pk}(\hat{n})&=[b_\nu-b_\zeta(\delta^2_1+\delta^2_2)]\Delta X(\hat{n})\, ,\\
        b_\nu&=\frac{1}{\sigma_0}\frac{\nu-\gamma\bar{u}}{1-\gamma^2}\label{eq:birefbnu}\, ,\\
        b_\zeta&=\frac{1}{\sigma_2}\frac{\bar{u}-\gamma\nu}{1-\gamma^2}\, ,\label{eq:birefbzeta}
    \end{align}
\end{subequations}
where $\Delta X$ is either the $E$ map, $\Delta E$, or $T$ map, $\Delta T$, giving the anisotropy amplitude for the peak intensity at $\hat{n}$, and $b_\nu$ and $b_\zeta$ are the peak bias parameters. We define the unitless peak height $\nu$,
\begin{align}
    \nu\equiv\Delta X/\sigma_0
\end{align}
for the peak height above the average fluctuation ($\sigma_0$), and the quantity 
\begin{align}
    \gamma\equiv\sigma_1^2/(\sigma_0\sigma_2)\, ,
\end{align}
which compares the peak height and its first and second derivative at the peak location. This can be calculated using the root mean square of the $j$th derivative of the $T$ or $E$ fluctuations,
\begin{align}
    \sigma_j^2=\frac{1}{4\pi}\sum_\ell (2\ell+1)[\ell(\ell+1)]^j C_\ell^{XX}(W_\ell^X)^2\, ,
\end{align}
where $C_\ell^{XX}$ is the power spectrum of the map from which we find the peaks ($T$- or $E$-mode map in this case, including additional effects such as the noise), and $W_\ell^X$ represents the beam, pixel window function and any other additional smoothing applied to the map before finding the peaks.
The mean curvature, $\bar{u}$, is given by 
\begin{subequations}
    \begin{align}
        \bar{u}&\equiv\frac{G_1(\gamma,\gamma\nu)}{G_0(\gamma,\gamma\nu)}\, ,\, \mathrm{with} \\
        G_n(\gamma,x_*)&\equiv\int_0^\infty dx\ x^nf(x)\frac{\exp\left[-\frac{(x-x_*)^2}{2(1-\gamma^2)}\right]}{\sqrt{2\pi(1-\gamma^2)}}\, ,\\
        \mathrm{and}\, f(x)&=x^2-1+\exp(-x^2)\, .
    \end{align}  
\end{subequations}
In previous stacking analyses the averaged bias parameters were used,
\begin{subequations}
    \begin{align}
        &\bar{n}_\mathrm{pk}(\nu_t)=\frac{\sigma_2^2}{(2\pi)^{3/2}(2\sigma_1^2)}\int^\infty_{\nu_t}d\nu\ e^{-\nu^2/2}G_0(\gamma,\gamma\nu)\, ,\\
        &\bar{b}_\nu=\frac{1}{\bar{n}_\mathrm{pk}(\nu_t)}\frac{\sigma_2^2}{(2\pi)^{3/2}(2\sigma_1^2)}\int^\infty_{\nu_t}d\nu\ e^{-\nu^2/2}G_0(\gamma,\gamma\nu)b_\nu\, ,\\
        &\bar{b}_\zeta=\frac{1}{\bar{n}_\mathrm{pk}(\nu_t)}\frac{\sigma_2^2}{(2\pi)^{3/2}(2\sigma_1^2)}\int^\infty_{\nu_t}d\nu\ e^{-\nu^2/2}G_0(\gamma,\gamma\nu)b_\zeta\, ,
    \end{align}  
\end{subequations}
for hot spots, with integration limits of $-\infty$ to $-|\nu_t|$ for cold spots, where $\nu_t$ is some threshold peak height. However, for our analysis, we choose to use the actual bias parameters (\cref{eq:birefbnu,eq:birefbzeta}) for each peak (which change depending on the peak's height), since this allows us to more optimally weight the different peaks. We also found that for different data splits this leads to more consistent results. When comparing all the data, the averaged bias parameters or the peak-by-peak bias parameters performed comparably.  Further discussion of this choice can be found in \cref{app:peakbiasparams}. 

We obtain the profile at a peak in the polarisation maps by calculating
\begin{subequations}
\label{equ:QrUrthe}
    \begin{align}
        \langle Q_r^X\rangle&=-\int \frac{\ell d\ell}{2\pi}W_\ell^X W_\ell^P(b_\nu+b_\zeta\ell^2)C_\ell^{XE}J_2(\ell\theta)\, ,\label{equ:QrUrthe1}\\
        \langle U_r^X\rangle&=-\int \frac{\ell d\ell}{2\pi}W_\ell^X W_\ell^P(b_\nu+b_\zeta\ell^2)C_\ell^{XB}J_2(\ell\theta)
        \, ,\label{equ:QrUrthe2}
    \end{align}  
\end{subequations}
\noindent where $J_2(\ell\theta)$ is the second-order Bessel function of the first kind, $\theta$ is a radial vector centred at a peak, and $X$ can be $E$ or $T$, representing the map where the peaks were found. These calculations determine the expected profile for a particular map's power spectrum. Note that the bias parameters should be calculated using the power spectrum derived from the map, so this should include all the noise and window function effects, whereas the $C_\ell^{XX}$ in \cref{equ:QrUrthe} should be the theory you would like to compare the data to, which is typically a noise-free ideal power spectrum without the effects of cosmic birefringence. 

If we substitute \cref{eq:cls_biref1,eq:cls_biref2,eq:cls_biref3,eq:cls_biref4,eq:cls_biref5,eq:cls_biref6} into \cref{equ:QrUrthe} we can obtain an estimate for the theoretical profiles, assuming some angle $\beta$,
\begin{subequations}
\label{eq:QrUrprofE_beta}
    \begin{align}
        \langle Q_r^E\rangle=-&\cos^2(2\beta)\int \frac{\ell d\ell}{2\pi}W_\ell^E W_\ell^P({b}_\nu+{b}_\zeta\ell^2)C_\ell^{EE}J_2(\ell\theta)\, ,\\
        \langle U_r^E\rangle=-&\frac{1}{2}\sin(4\beta)\int \frac{\ell d\ell}{2\pi}W_\ell^E W_\ell^P({b}_\nu+{b}_\zeta\ell^2)C_\ell^{EE}J_2(\ell\theta)\, ,
    \end{align}  
\end{subequations}
\noindent for $E$ peaks, and 
\begin{subequations}
\label{eq:QrUrprofT_beta}
    \begin{align}
        \langle Q_r^T\rangle&=-\cos(2\beta)\int \frac{\ell d\ell}{2\pi}W_\ell^T W_\ell^P({b}_\nu+{b}_\zeta\ell^2)C_\ell^{TE}J_2(\ell\theta)\, ,\\
        \langle U_r^T\rangle&=-\sin(2\beta)\int \frac{\ell d\ell}{2\pi}W_\ell^T W_\ell^P({b}_\nu+{b}_\zeta\ell^2)C_\ell^{TE}J_2(\ell\theta)\, ,
    \end{align}  
\end{subequations}
\noindent for $T$ peaks, where we assume that the $C_\ell^{BB}$ are zero. 
\begin{figure}[tbp!]
    \centering
    \includegraphics[width=0.65\linewidth]{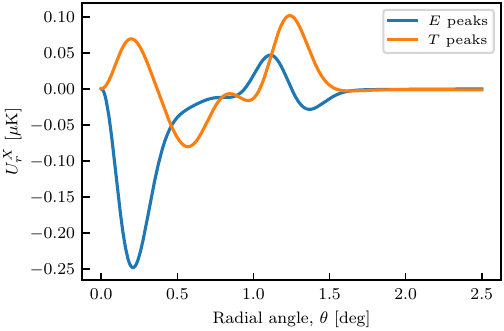}
    \caption{Expected profiles for $U_r^X$, $X\in T,E$, assuming a birefringence angle of \ang{0.3} and the average bias parameters $b_\nu$ and $b_\zeta$. For no birefringence angle we expect the profile to be zero. 
    }
    \label{fig:birefprofiles}
\end{figure}
We take the angular distance $\theta$, out to \ang{2.5} from the peak centre, since the profile converges to zero at these distances (see \cref{fig:birefprofiles}). 

To find the extrema (hot and cold spots) in the $T$ and $E$ maps we use the nearest-neighbour method, considering the nearest eight pixels (where if the nearest eight pixels are larger/smaller than the central pixel it is taken as a peak). We tested our results using the Hessian to find the critical points \cite{2019Jow}, but this did not outperform the nearest-neighbour method and introduced additional complexity. 
We choose to only keep the hot spots that are positive, and cold spots that are negative (a threshold peak height $\nu$ of zero i.e.\ any local positive peaks with a negative height were excluded). We tested other thresholds and found that keeping all the peaks (i.e.\ a lower threshold) led to higher noise in the averaged results, due to a higher number of `noise' peaks entering the analysis, whereas using a higher threshold did not significantly improve the analysis. For an extreme enough cut the final significance of the results was reduced due to the small number of analysed peaks. This also follows the conventions chosen in Refs.~\cite{planck2014-a23,2017Contreras}.
In addition, to reduce the presence of noise peaks, we choose to smooth the $T$ and $E$ maps with an extra \ang{;10;} FWHM Gaussian beam, giving them a final effective beam of around \ang{;11.2;}. We also cut all $\ell$s above 1500, since information on smaller scales is noise-dominated, and the majority of the power for estimating $\beta$ comes from $\ell<1500$. Other $\ell$ cuts were tested and did not significantly change the results, (further justification of this choice can be found in \cref{app:ellcut}). 
Any peaks within \ang{2.5} of a masked region are also discarded, since the full profile could not be recovered for the masked peaks and they add marginal signal (while amplifying the noise). 
We find on the order of $1.9\times10^5$ peaks in an $E$ map, and $6.6\times10^4$ peaks in a $T$ map. Since each peak (out to a radius of \ang{2.5}) covers around 20 square degrees on the sky there is necessarily some overlap between the peaks.

For each peak a linear least squares estimate for the best-fit birefringence angle is obtained, as well as its goodness of fit. In this way, we can favour peaks that have a better fit compared to those with a poorer fit. 
For each peak, $m$, we calculate a best-fit angle, $\beta_m$ using,
    \begin{align}
        \beta_m&=\frac{1}{2}\frac{\sum_p^N \hat{U}_r(\theta_p)\tilde{U}_r(\theta_p)}{\sum_p^N\tilde{U}_r(\theta_p)\tilde{U}_r(\theta_p)}\, ,\\
        \sigma_{\beta_m}^2&=\frac{\sum_p^N(\hat{U}_r(\theta_p)-2\beta_m\tilde{U}_r(\theta_p))^2}{(N-1)\sum_p^N\tilde{U}_r(\theta_p)\tilde{U}_r(\theta_p)}\, ,
    \label{eq:betabestfit}
    \end{align}
where $\tilde{U}_r(\theta_p)$ is the theory profile and $\hat{U}_r(\theta_p)$ is the value of the data, $p$ is each pixel, $N$ is the total number of pixels within $\theta=\ang{2.5}$ of the peak. We can average those for the unmasked sky using
    \begin{align}
        \langle \beta \rangle=\frac{\sum_m \beta_m/\sigma_{\beta_m}^2}{\sum_m 1/\sigma_{\beta_m}^2}\, .
    \end{align}

We can obtain the uncertainty on $\langle \beta \rangle $ either using the variation in the simulation results or using the variance of the weighted mean,
\begin{align}
    \langle \sigma_\beta^2 \rangle=\left(\frac{1}{\sum_m(1/\sigma_{\beta_m}^2)}\right)^2\sum_m (1/\sigma_{\beta_m}^2)^2(\langle \beta \rangle-\beta_m)^2\ .
    \label{eq:biref_sigmab}
\end{align}

Starting in \cref{eq:betabestfit}, we have assumed that the pixel-to-pixel covariance around a particular peak is the same for each pixel and is diagonal. We could instead scale by the hits map or some other metric, but since the pixels are all relatively close, we expect them to have similar noise levels. We are also ignoring correlations between the pixels, since these are expected to be small. Future work could aim to implement the full pixel correlations into the analysis and potentially do the analysis at a higher resolution than was done here.

The simulation uncertainties are derived from the width of the histograms of 300 simulations; however, data uncertainties are calculated using \cref{eq:biref_sigmab}. This allows us to account for the different noise in each component-separation method. From the simulations, we find that the histogram width and $\langle\sigma^2_\beta\rangle$ are equivalent measures.

It is simple to extend the analysis to different masks, since we have already calculated a separate estimate for each peak. For each mask we average the peaks that are unmasked in the same way as described for the full sky; this means that estimates for different masked areas are of course correlated with one another if they have overlap, and also that larger masks will naturally increase the error because less data are being averaged.

\section{Results}
\label{sec:res}
\subsection{Full sky results}
In this work, we do not recalibrate the miscalibration angles using the Galactic foreground. Therefore our values of $\beta$ are the sum of the miscalibration and birefringence angles. 
We find consistent values compared with the previously published results for the whole sky, with an average for the common mask $\beta=\ang{0.46}\pm\ang{0.04}(\mathrm{stat.})\pm\ang{0.28}(\mathrm{syst.})$ for \SEVEM\ and $\beta=\ang{0.48}\pm\ang{0.04}(\mathrm{stat.})\pm\ang{0.28}(\mathrm{syst.})$ for \Commander\ (see \cref{tab:constraints}). The systematic uncertainty, $\pm\ang{0.28}(\mathrm{syst.})$,
comes from the uncertainty in the polarimeter miscalibration from \planck. This result is derived from ground measurements \cite{rosset2010} and in-flight calibration using the Crab Nebula as the primary calibration source \cite{planck2014-a09}. The results are mostly constrained by the $E$ peaks, with a weaker constraint from the $T$ peaks. This is expected since if we consider \cref{eq:cls_biref5,eq:cls_biref6}, then 
\begin{align}
    \frac{\partial C_\ell^{TB}}{\partial \beta}=2\cos(2\beta)C_\ell^{TE}\approx 2 C_\ell^{TE}\\
    \frac{\partial C_\ell^{EB}}{\partial \beta}=2\cos(4\beta)\left(C_\ell^{EE}-C_\ell^{BB}\right)\approx 2C_\ell^{EE}\, ,
\end{align}
for $\beta\ll 1$ and $C_\ell^{BB}\ll C_\ell^{EE}$. The Fisher information matrix gives us, 
\begin{align}
    \sigma_\beta=\sqrt{F_{\beta\beta}^{-1}}\, ,
\end{align}
with 
\begin{align}
    F_{\beta\beta}^{TB}\propto\frac{\partial C_\ell^{TB}}{\partial\beta}\frac{1}{\mathrm{Var}[C_\ell^{TB}]}\frac{\partial C_\ell^{TB}}{\partial\beta}\approx \frac{4\left(C_\ell^{TE}\right)^2}{C_\ell^{TT}C_\ell^{BB}}\, ,
\end{align}
for $TB$ and,
\begin{align}
    F_{\beta\beta}^{EB}\propto\frac{\partial C_\ell^{EB}}{\partial\beta}\frac{1}{\mathrm{Var}[C_\ell^{EB}]}\frac{\partial C_\ell^{EB}}{\partial\beta}\approx\frac{4C_\ell^{EE}}{C_\ell^{BB}}\, ,
\end{align}
for $EB$. From here we can compare the inverse of the two results and note that $C_\ell^{TT}/\left(C_\ell^{TE}\right)^2$ is much larger than $1/C_\ell^{EE}$, and as expected the variance on $\beta$ from $E$ peaks will be much less than $T$ peaks. We have taken several liberties here and neglected noise, but it suffices as an approximation. For a similar analysis see Ref.~\cite{2022Abghari}. 

\begin{figure}[tbp!]
    \centering
    \includegraphics[width=0.65\hsize]{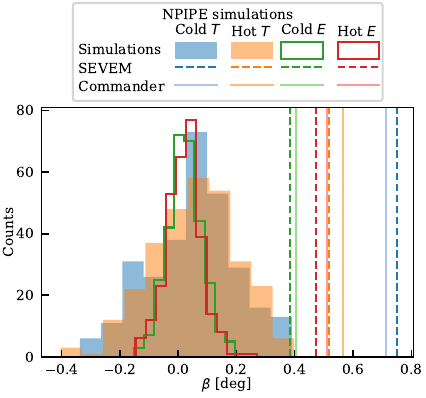}
    \caption{Birefringence results for each data cut. Histograms are from the 300 end-to-end NPIPE simulations with no birefringence component. We can see that the true data are inconsistent with a birefringence of zero at more than the $2\sigma$ level for each data cut. Note, however, that the systematic error (of \ang{0.28}) is not included here, which dominates the error budget.
    }
    \label{fig:commonMask_birefresults}
\end{figure}

The advantage of real space analysis is the ease with which we can check for spatially-dependent systematic effects. For example, we expect that there should be consistency between the different extrema splits (hot and cold extrema, and $E$ and $T$ peaks) for pure cosmic birefringence, and we find that for the common mask analysis, the different splits are consistent with one another (see \cref{fig:commonMask_birefresults}). The $E$ and $T$ peaks are likely to be affected by different systematic effects, however, so differences between these splits might point to lingering systematic effects in the data.  We can also perform a more detailed isotropy analysis as shown in \cref{ssec:noniso}, to see if any areas of the sky are contributing more or less signal. Significant variation across the sky would indicate that foreground systematic effects could be playing a role. Obvious masking strategies are to use masks that remove areas with significant dust or synchrotron, and we also consider a north versus south mask cut. 

\subsection{Comparing extrema}
\begin{figure}[tbp!]
    \centering
    \includegraphics[width=0.65\hsize]{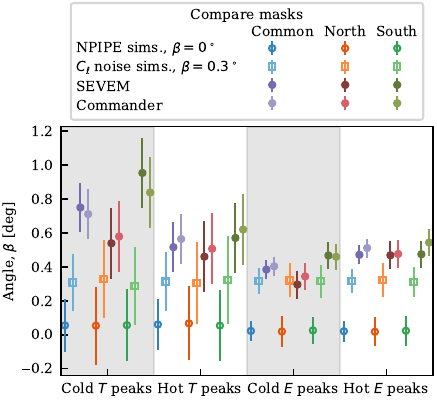}
    \caption{Results from the end-to-end NPIPE simulations (with $\beta=\ang{0}$), simulations with noise matching the \SEVEM\ results (though no realistic foregrounds, and $\beta=\ang{0.3}$), results from the \SEVEM\ PR4 results and from the \Commander\ PR4 results for the North and South mask (North meaning the northern hemisphere is removed). The simulation points are the average from all simulations, and the error bar is the $1\sigma$ standard deviation from the simulations. Due to the large mask the uncertainty is correspondingly higher; however, it does appear that there is a trend for the masked southern hemisphere results to be higher for each data cut. There appears to be a slight high bias in the results for the end-to-end simulations. This disappears for cosmic variance limited results and results with simple white noise (see \cref{app:CVL_biref}), so this may be due to the inclusion of realistic telescope effects in the end-to-end results, and in any case is much smaller than the $1\sigma$ error bars. }
    \label{fig:biref_northsouthscatter}
\end{figure}

\begin{figure}[tbp!]
    \centering
    \includegraphics[width=0.65\hsize]{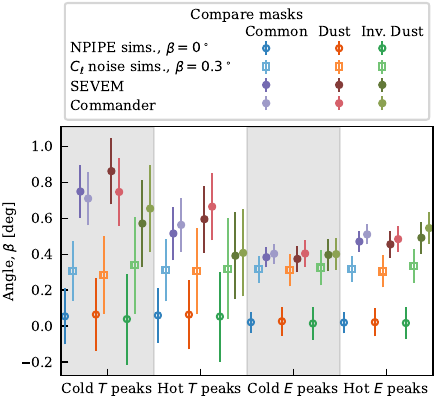}
    \caption{Similar to \cref{fig:biref_northsouthscatter}, but for the dust and inverse dust masks. Recall that the dust mask should remove more of the dust-contaminated regions, whereas the inverse dust mask leaves the dustier regions. Here we see smaller birefringence in the dustier areas in the $T$ peaks, this will be in stark contrast to the synchrotron peaks in \cref{fig:biref_syncscatter}.}
    \label{fig:biref_dustscatter}
\end{figure}

\begin{figure}[tbp!]
    \centering
    \includegraphics[width=0.65\hsize]{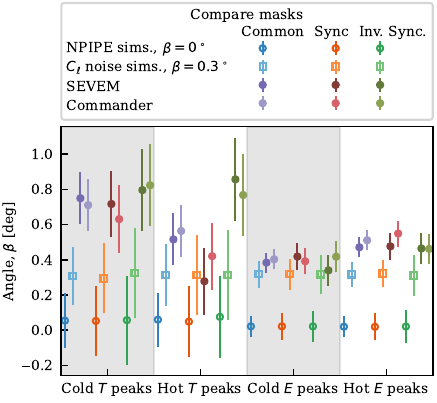}
    \caption{Similar to \cref{fig:biref_northsouthscatter,fig:biref_dustscatter}, but for the synchrotron and inverse synchrotron masks. Recall that the synchrotron mask removes areas of high synchrotron, and inverse synchrotron keeps the areas of high synchrotron. Here we can see that the temperature peaks seem to show some shifts, with areas of high synchrotron leading to higher central values, particularly for the hot $T$ peaks. }
    \label{fig:biref_syncscatter}
\end{figure}

The four kinds of extrema (hot and cold, $T$ and $E$) might initially appear to be consistent with one another (in particular when including the uncertainty from the miscalibration angle), however it is important to note that the four should be affected in the same way by both miscalibration and cosmic birefringence, thus any discrepancy between them to within the statistical error at least, is most easily attributed to foregrounds, which will affect all four measures differently. $T$ and $E$ may also be discrepant, since they will be affected by different systematic effects. In our case, the largest offset is with the cold temperature peaks (see \cref{tab:constraints}), though the significance is not large enough to claim a detection of foreground $TB$.
\begin{table}[tbp!]
\begingroup
\newdimen\tblskip \tblskip=5pt
\nointerlineskip
\vskip -3mm
\footnotesize
\setbox\tablebox=\vbox{
   \newdimen\digitwidth
   \setbox0=\hbox{\rm 0}
   \digitwidth=\wd0
   \catcode`*=\active
   \def*{\kern\digitwidth}
   \newdimen\signwidth
   \setbox0=\hbox{+}
   \signwidth=\wd0
   \catcode`!=\active
   \def!{\kern\signwidth}
\halign{\tabskip 0pt \hbox to 0.8in{#\leaderfil}\tabskip 0pt&
         \hfil#\hfil\tabskip 10pt&
         \hfil#\hfil\tabskip 10pt&
         \hfil#\hfil\tabskip 10pt&
         \hfil#\hfil\tabskip 10pt&
         \hfil#\hfil\tabskip 10pt&
         \hfil#\hfil\tabskip 10pt&
         \hfil#\hfil\tabskip 0pt\cr                           %
\noalign{\doubleline\vskip 1pt}
\omit\hfil & !Cold $T$& !Hot $T$ & All $T$  & !Cold $E$ & !Hot $E$ & !All $E$ & Combined \cr  
\omit\hfil Method\hfil& [deg]& [deg]& [deg] & [deg]& [deg]& [deg]& [deg]\cr 
\noalign{\vskip 4pt\hrule\vskip 6pt}
\noalign{\vskip 3pt}
{\tt \commander}& 
$ 0.71 \pm 0.15 $&$ 0.56 \pm 0.15 $&  $ 0.64 \pm 0.11 $& $ 0.40 \pm 0.06 $& $ 0.51 \pm 0.06 $& $ 0.46 \pm 0.04 $& $ 0.48 \pm 0.04 $\cr
\noalign{\vskip 3pt}
{\tt \sevem}& 
$ 0.75 \pm 0.15 $& $ 0.52 \pm 0.15 $&  $ 0.63 \pm 0.10 $& $ 0.38 \pm 0.06 $& $ 0.47 \pm 0.06 $& $ 0.43 \pm 0.04 $& $ 0.46 \pm 0.04 $\cr
\noalign{\vskip 3pt\hrule\vskip 4pt}}}
\endPlancktable                   
\endgroup
\caption{Mean values and ($1\,\sigma$) statistical uncertainties for $\beta$ (in degrees)
derived from the stacking analysis for all component-separation methods, coming
from hot spots, cold spots, and all extrema.}
\label{tab:constraints}
\end{table}
There is an interesting trend to note when we consider the different masks (see \cref{fig:biref_northsouthscatter,fig:biref_dustscatter,fig:biref_syncscatter}). Across all the masks we see fairly robust results from the $E$ peaks, the largest difference being a small shift in the cold $E$ peaks between the North and the South but less than $1\sigma$. However, for the $T$ peaks, we see much greater variability, results from the northern hemisphere (South mask) have higher $\beta$ values than the southern hemisphere, by about 1$\sigma$. 
Similarly, the dustier regions (from the inverse dust mask) show lower $\beta$ for both \sevem\ and \commander\ in the $T$ peaks, and the higher synchrotron regions show higher $\beta$. This is perhaps expected, after noting that the northern hemisphere is the location of the North Galactic Spur, with potentially higher synchrotron variations. The power asymmetry between the two hemispheres could also be a contributor \cite{planck2016-l07}. 
The result is subtle and only present in the $T$ peaks, so we cannot draw any strong conclusions about the foreground production of $EB$ or $TB$, but we do recommend further investigation before attributing any cosmological origin to a detection of $\beta$.  

\subsection{Anisotropic birefringence}
\label{ssec:noniso}
As well as the map-space method being useful for investigating anisotropic systematic effects, it can also be used to look for an anisotropic birefringence signal.
Anisotropic birefringence is of particular interest due to its potential cosmological implications, especially on large angular scales. There is a wide range of possibilities for angular variation of anisotopic birefringence.  Here, we are particularly interested in the large-scale variations on the sky, since Ref.~\cite{2017Contreras} found a larger-than-expected dipole term and we want to follow up on those findings. Anisotropic birefringence is also resilient against the polarimeter miscalibration angle, since that should be constant across the sky. 

\begin{figure}[tbp!]
    \centering
    \includegraphics[width=0.65\hsize]{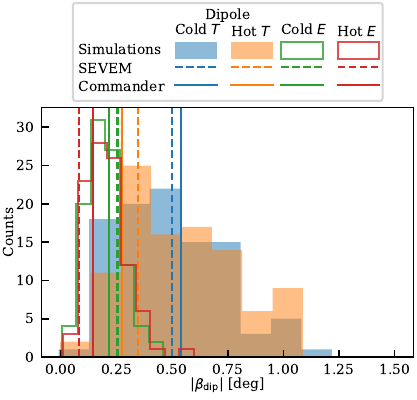}
    \caption{Test for anisotropic birefringence considering a birefringence dipole, $\beta_\mathrm{dip}$. Using a modified \texttt{fit\_dipole} functionality from \texttt{healpy} with the peaks found in the previous analysis, we can see that both the \SEVEM\ and \Commander\ maps have results consistent with the simulations (i.e.\ consistent with zero dipole in the birefringence signal across the sky).}
    \label{fig:birefdipole}
\end{figure}

For this purpose, we use a modified version of the \texttt{fit\_dipole} functionality from \texttt{healpy} to measure the dipole magnitude in the weighted birefringence signal across the sky, $\beta_\mathrm{dip}$. We weigh the peaks with their variance and perform a least-squares fit to a dipole vector across the sky. The fitting minimises the sum of squared residuals 
$\sum_i \frac{(y_i-\beta_\mathrm{mon} - \vec{x}_i \cdot \vec{\beta}_{\mathrm{dip},i})^2}{\sigma^2_i}$ or in matrix notation,
\begin{align}
S= (\mathbf{Y-X\beta})^{\sf T} \mathbf{W} (\mathbf{Y-X\beta})
\end{align}
where $\mathbf{Y}$ is a vector of birefringence angle values for peaks, $\mathbf{X}$ is a matrix containing the coordinates of the peaks and $\mathbf{W}$ is the weight matrix given by $\mathbf{W} = \text{diag}(1/ \sigma_i^2)$. $\beta$ represents the coefficients to be fitted, monopole and dipole terms of birefringence in this case. The fitted values for $\mathbf{\beta}$ are calculated using
\begin{align}
\hat{\mathbf{\beta}} = \mathbf{(X^{\sf T} W X)^{-1}X^{\sf T}WY}.
\end{align}
Although the least-squares method is biased when a mask is applied, our mask is sufficiently small that the bias is negligible.  
We find the results from the \sevem\ and \commander\ are consistent with the simulations with zero birefringence dipole signal (see \cref{fig:birefdipole}). 

\section{Conclusions}
\label{sec:birefconc}
Analysis of the PR4 data release from \planck, using a map-space approach, yields values of $\beta$ that are consistent compared to previous results, with $\beta=\ang{0.46}\pm\ang{0.04}(\mathrm{stat.})\pm\ang{0.28}(\mathrm{syst.})$ for \SEVEM\ CMB maps and $\beta=\ang{0.48}\pm\ang{0.04}(\mathrm{stat.})\pm\ang{0.28}(\mathrm{syst.})$ for \Commander\ CMB maps. This method (at least in the simplest form described here) does not account for the miscalibration offset in the polarimeters and so cannot distinguish systematic effects from the polarimeters calibration from any cosmological signal.
Tests for different masks indicate a preference for somewhat higher angle measures in the northern hemisphere (South mask) compared to the southern hemisphere (North mask). Different data cuts (hot and cold spots from $E$ and $T$ peaks) have some inconsistency with one another, the $T$ peaks showing a preference for slightly higher values of $\beta$, but also with a larger uncertainty (since the $T$ peaks have less containing power on birefringence measurements). This cannot be attributed to cosmic birefringence or a miscalibration angle offset, which would alter all data cuts equally, so could be a hint of a foreground $TB$ signal or other systematic effects.

We also find that the temperature peaks are more affected by masks correlated with stronger dust or synchrotron foregrounds. Results from Ref.~\cite{planck2014-XXX} show that the dust polarisation has some $TB$ signal and modelling studies suggest that there could also be Galactic $EB$ \cite{2021Clark_galPol,2023MagDust,2024magdust2}.
Our analysis suggests that Galactic dust $TB$ and $EB$ might affect the birefringence measure, but synchrotron may also have a non-negligible effect. The results for temperature peaks are higher for regions with more synchrotron than those with less. Since we would expect the various data cuts to match one another, then it does seem that a cleaner data set will be required to further probe cosmic birefringence. 

The map-space analysis method is a valuable tool to use to search for inconsistencies in the data, especially those that vary spatially and with different data cuts. It should thus be used in conjunction with harmonic space analyses, which could suffer from different systematic effects. For clean maps, without residual foregrounds (for example in the $C_\ell$ noise simulations), the method finds very consistent results for an input birefringence angle of \ang{0.3}.

We find no significant dipole detection in the birefringence signal. Future analysis with new data sets, such as from CMB-S4~\cite{2022ApJCMBs4}, Simons Observatory~\cite{2023SO1,2017Polocalc,2024SO2,2024SO3,2021SO4} and \textit{LiteBIRD}~\cite{2023PTEPLiteBIRD,2025LiteBIRDBiref}, will give independent data sets with better control of the foregrounds and instrumental polarisation angles. These will hopefully help break the degeneracy between the miscalibration angle offsets and any birefringence and may be able to give tighter constraints on any anisotropic birefringence signals. 

\begin{acknowledgments}
We would like to thank Eiichiro Komatsu, Belen Barreiro, Yuto Minami and Matthieu Tristram for useful comments. Some of the results in this paper have been derived using the \texttt{healpy}~\footnote{\url{https://github.com/healpy/healpy}}  and \texttt{HEALPix}~\footnote{\url{https://healpix.sourceforge.io}}  packages.
This work made use of \texttt{astropy},\footnote{\url{https://www.astropy.org}} a community-developed core Python package and an ecosystem of tools and resources for astronomy, as well as \texttt{numpy}~\footnote{\url{https://numpy.org/}} and \texttt{matplotlib}.\footnote{\url{https://matplotlib.org}} We acknowledge the support of the Natural Sciences and Engineering Research Council of Canada (NSERC).
LTH was supported by a Killam Postdoctoral Fellowship and a CITA National Fellowship. 
\end{acknowledgments}

\appendix

\section{Cosmic-variance limited tests}
\label{app:CVL_biref}
\begin{figure}[tbp!]
    \centering
    \includegraphics[width=0.65\hsize]{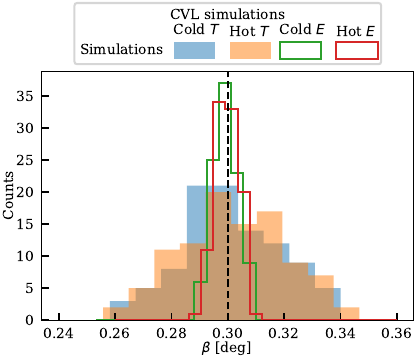}
    \caption{Cosmic-variance-limited (noise-free) tests. This represents the best possible outcome with a birefringence angle of $\ang{0.3}$, still including a mask and beam, and at a resolution of $N_\mathrm{side}=2048$ (see also \cref{tab:CVLtests}). }
    \label{fig:biref_cvlsimshist}
\end{figure}

\begin{figure}[tbp!]
    \centering
    \includegraphics[width=0.65\hsize]{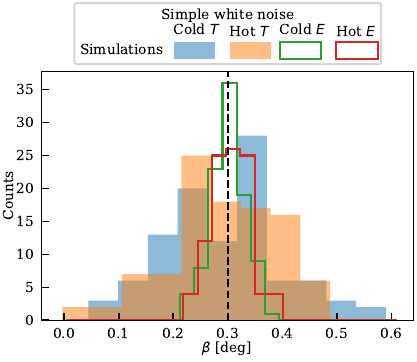}
    \caption{Flat white noise tests. These simulations were made with flat (in $\ell$ space) white noise of \SI{12.6}{\micro\K} per pixel in the temperature maps, and \SI{18.3}{\micro\K} in the polarisation $Q$ and $U$ maps. This should approximate the results for near-perfect component separation, leaving only a small level of perfectly white noise. }
    \label{fig:biref_wnsimshist}
\end{figure}

We run some tests using our new pipeline to determine the best possible results achievable. These are done with 100 noise-free simulated maps with a birefringence angle of \ang{0.3}, but still including a beam of \ang{;5;}, at a resolution of $N_\mathrm{side}$=2048, making the same $\ell_{\mathrm{max}}$ cut of 1500, and using the same peak thresholds. We also still include the common mask. Thus this primarily tests the effect of the noise on the data (see \cref{fig:biref_cvlsimshist}). We additionally show results for 100 noisy simulations, with a simple white noise level of \SI{12.6}{\micro\K} per pixel in the temperature maps, and \SI{18.3}{\micro\K} in the polarisation $Q$ and $U$ maps (these numbers were chosen based on the standard deviation of the histogram of the NPIPE noise maps), see \cref{fig:biref_wnsimshist}. This is a confirmation of the method since the pipeline would detect a signal of $\ang{0.300}\pm\ang{0.003}$ for cosmic variance limited simulations, and at $\ang{0.302}\pm\ang{0.022}$ for the simplistic white noise simulations (a more realistic though still very high detection level). These are in contrast to the results from the NPIPE-like noise simulations, at $\ang{0.32}\pm\ang{0.05}$, and finally the NPIPE simulations (now with no birefringence angle), at $\ang{0.03}\pm\ang{0.04}$. 
For other tests, \cref{tab:CVLtests} shows results for lower resolution maps with and without beam, pixel window function and noise (true cosmic variance with perfect resolution), as well as results with different thresholds and lower $\ell_\mathrm{max}$. All show good recovery of the input birefringence angle.

\begin{table*}[tbp!]
\begingroup
\newdimen\tblskip \tblskip=5pt
\nointerlineskip
\vskip -3mm
\footnotesize
\setbox\tablebox=\vbox{
   \newdimen\digitwidth
   \setbox0=\hbox{\rm 0}
   \digitwidth=\wd0

   \newdimen\signwidth
   \setbox0=\hbox{+}
   \signwidth=\wd0
   \catcode`!=\active
   \def!{\kern\signwidth}
\halign{\tabskip 0pt \hbox to 1.5in{#}\tabskip 0pt&
         \hfil#\hfil\tabskip 12pt&
         \hfil#\hfil\tabskip 12pt&
         \hfil#\hfil\tabskip 12pt&
         \hfil#\hfil\tabskip 12pt&
         \hfil#\hfil\tabskip 12pt&
         \hfil#\hfil\tabskip 0pt\cr                           
\noalign{\doubleline\vskip 1pt}
\centering

   Simulation test\hfil &$N_\mathrm{side}$&c$T$ [deg]&h$T$ [deg]&c$E$ [deg]&h$E$ [deg]&All [deg]\cr
\noalign{\vskip 4pt\hrule\vskip 6pt}
\multirow{3}{*}{Without $N_\ell$, $B_\ell$ or $P_\ell$\hfil}
&64 & $ 0.30 \pm 0.08 $ & $ 0.29 \pm 0.06 $ & $ 0.30 \pm 0.03 $ & $ 0.30 \pm 0.03 $ & $ 0.30 \pm 0.02 $\cr
   &128 &$ 0.31 \pm 0.05 $ & $ 0.32 \pm 0.06 $ & $ 0.32 \pm 0.03 $ & $ 0.32 \pm 0.03 $ & $ 0.32 \pm 0.02 $\cr
   &512 &$ 0.33 \pm 0.02 $ & $ 0.33 \pm 0.02 $ & $ 0.30 \pm 0.01 $ & $ 0.31 \pm 0.01 $ & $ 0.31 \pm 0.01$\cr
\noalign{\vskip 3pt\hrule\vskip 4pt}
\multirow{3}{*}{\shortstack{With $N_\ell$, $B_\ell$ and $P_\ell$}}
&64&$ 0.31 \pm 0.14 $ & $ 0.31 \pm 0.12 $ & $ 0.30 \pm 0.05 $ & $ 0.29 \pm 0.05 $ & $ 0.30 \pm 0.03 $\cr
&128&$ 0.30 \pm 0.07 $ & $ 0.30 \pm 0.05 $ & $ 0.31 \pm 0.04 $ & $ 0.32 \pm 0.04 $ & $ 0.31 \pm 0.02 $\cr
&512&$ 0.30 \pm 0.04 $ & $ 0.29 \pm 0.03 $ & $ 0.36 \pm 0.02 $ & $ 0.36 \pm 0.01 $ & $ 0.35 \pm 0.01$\cr
\noalign{\vskip 3pt\hrule\vskip 4pt}
$\ell_\mathrm{max}<1000$\hfil &512&$ 0.30 \pm 0.04 $ & $ 0.30 \pm 0.04 $ & $ 0.31 \pm 0.01 $ & $ 0.31 \pm 0.01 $ & $ 0.31 \pm 0.01 $\cr
\noalign{\vskip 3pt\hrule\vskip 4pt}
Threshold, $\nu=\phantom{-}1$\hfil &128&$ 0.31 \pm 0.07 $ & $ 0.30 \pm 0.07 $ & $ 0.31 \pm 0.03 $ & $ 0.30 \pm 0.04 $ & $ 0.30 \pm 0.02$\cr
Threshold, $\nu=\phantom{-}3$\hfil &128&$ 0.26 \pm 0.30 $ & $ 0.34 \pm 0.25 $ & $ 0.32 \pm 0.07 $ & $ 0.31 \pm 0.07 $ & $ 0.32 \pm 0.05 $\cr
Threshold, $\nu=-5$ \hfil&128&$ 0.32 \pm 0.08 $ & $ 0.33 \pm 0.07 $ & $ 0.34 \pm 0.05 $ & $ 0.33 \pm 0.06 $ & $ 0.33 \pm 0.03 $\cr
\noalign{\vskip 3pt\hrule\vskip 4pt}}}
\endPlancktable                   
\endgroup
\caption{Cosmic-variance-limited results for different \nside, results with beam $B_\ell$ and pixel window function $P_\ell$, results with noise $N_\ell$ and beam and pixel window function, compared with results with only low $\ell$, and results for different peak thresholds (threshold of zero unless otherwise stated). Here `h' and `c' are for the hot and cold peaks, respectively (see also \cref{fig:biref_cvlsimshist}).}
\label{tab:CVLtests}
\end{table*}

\section{Peak bias parameters}
\label{app:peakbiasparams}
\begin{figure*}[tbp!]
    \centering
    \begin{subfigure}{0.65\textwidth}
        \includegraphics[width=\textwidth]{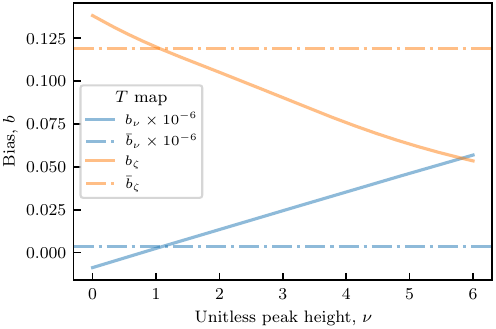}
    \end{subfigure}
    \begin{subfigure}{0.65\textwidth}
        \includegraphics[width=\textwidth]{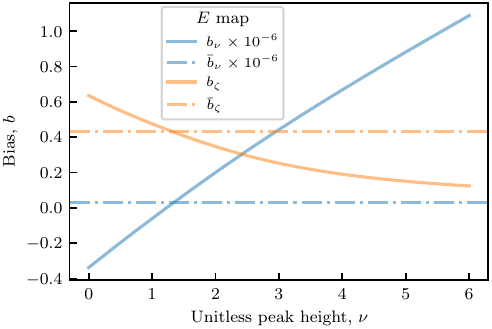}
    \end{subfigure}
    \caption{Investigation of peak bias. In previous studies the map-averaged bias parameters were assumed for all peaks (independent of the peak height), denoted here with the horizontal lines labelled $\bar{b}$. Here we use the individual bias parameter appropriate for each peak's height. Most peaks will be close to the crossing point of the $\bar{b}$ and $b$, near a threshold peak height $\nu$ of between 1 and 2.  
    Results from both \commander\ and \sevem\ are imperceptibly different, so only \sevem\ is shown here.}
    \label{fig:biref_biasparams}
\end{figure*}

We now want to check our approach to weighting the peaks. We calculate the bias parameters for each peak in this analysis, contrary to the method taken in Refs.~\cite{2017Contreras, komatsu2010}, where the average map bias parameters were used. 
The bias parameters are a function of the unitless peak height $\nu$, which depends on the amplitude of the peak above the r.m.s. temperature fluctuations, so more extreme peaks will have a greater dependence on the linear bias ($b_\nu$) than shallower peaks and will have a smaller dependence on the scale-dependent bias ($b_\zeta)$ than the shallower peaks (see \cref{fig:biref_biasparams}). In either case, the scale-dependent bias contains a factor of $\ell^2$ compared to the linear bias; so even though it is lower than the linear bias for low-multipoles, it will begin to dominate above $\ell\approx1000$.  The dot-dashed line in \cref{fig:biref_biasparams} clearly shows that using the average bias parameters is only the correct approximation for peak heights close to one. While peaks close to one are the most common, the results can be considerably different for the lowest and highest peaks. Future work could investigate the split of results between the highest peaks and the lowest peaks.  

\begin{figure}[tb]
    \centering
    \includegraphics[width=0.65\linewidth]{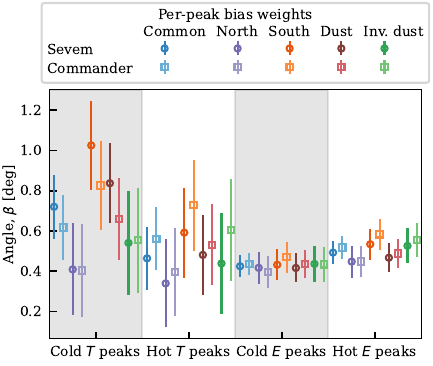}
    \caption{Test of the effect of weighting with individual peak bias. These results employ the per-peak bias parameter calculation, correctly weighting each peak based on its height. Results for different masks are very consistent, particularly when considering the $E$ peaks and compared with \cref{fig:birefnotppcomp}.}
    \label{fig:birefppcomp}
    \includegraphics[width=0.65\linewidth]{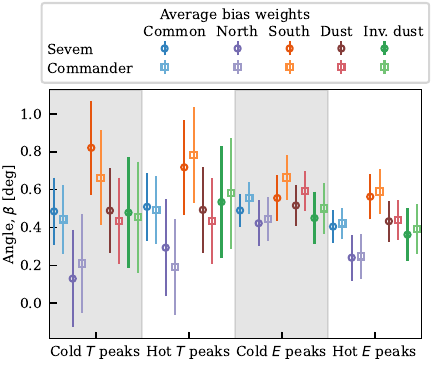}
    \caption{Test of the effect of weighting with average bias. These results employ the map average bias parameter calculation, weighting each peak based on the same full map average. Results for different masks vary, particularly when considering the $E$ peaks and comparing them to \cref{fig:birefppcomp}. }
    \label{fig:birefnotppcomp}
\end{figure}

This peak-weighting procedure does not have a large effect on the average (isotropic) birefringence calculation, but is done to ensure that inaccurate peak-by-peak bias calculations do not change the anisotropic measure. \Cref{fig:birefppcomp} shows the analysis of the data taking first the peak-by-peak bias parameters, customised for each peak's height, whereas \cref{fig:birefnotppcomp} shows the results using instead just the map-averaged bias parameter. Notably for different masks we see that the peak-by-peak bias parameters fluctuate less, indicating that some weightings of peaks in masked regions are more affected by the map average. The effect is subtle, but it is worth checking both analyses to ensure this choice does not have a large effect. 

\section{Justification for \texorpdfstring{$\ell_{\mathrm{max}}$}{l max} cut}
\label{app:ellcut}
\begin{figure*}[tbp!]
    \centering
    \begin{subfigure}{0.49\textwidth}
        \includegraphics[width=\textwidth]{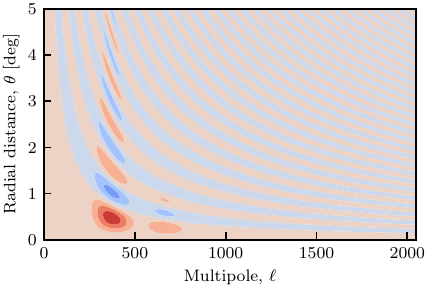}
        \caption{Contour for \cref{eq:lmaxA}. }
    \end{subfigure}
    \begin{subfigure}{0.49\textwidth}
        \includegraphics[width=\textwidth]{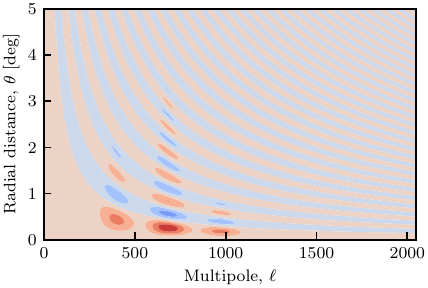}
        \caption{Contour for \cref{eq:lmaxB}. }
    \end{subfigure}
    \begin{subfigure}{0.49\textwidth}
        \includegraphics[width=\textwidth]{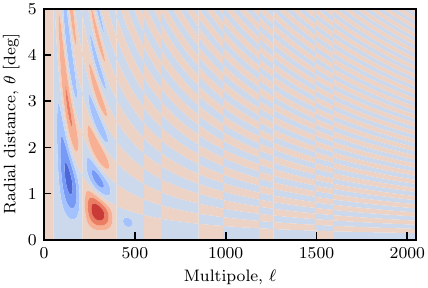}
        \caption{Contour for \cref{eq:lmaxC}. }
    \end{subfigure}
    \begin{subfigure}{0.49\textwidth}
        \includegraphics[width=\textwidth]{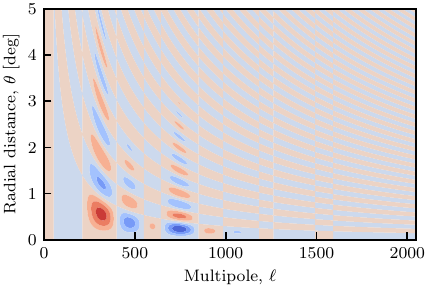}
        \caption{Contour for \cref{eq:lmaxD}. }
    \end{subfigure}
    \caption{Contour plots showing the power in the various terms in \cref{eq:QrUrterms}. The actual amplitude of each will vary depending on the magnitude of the bias parameters, but this figure serves to justify the minimum and maximum ranges of $\theta$ and $\ell$ relevant for the analysis. Note that due to the oscillatory nature of the functions, it is not always immediately obvious how much power will remain after summing along $\ell$ or $\theta$; for this reason, it may appear that significant power is still present for higher $\theta$ values, but when the summed profile is considered there is negligible power found above $\theta=\ang{2.5}$, as already seen in \cref{fig:birefprofiles}. }
    \label{fig:birefcontours}
\end{figure*}

We now return in more detail to the question of what multipole range to include by considering the theoretical functions $Q_r$ and $U_r$. 
In \cref{fig:birefcontours} we plot \cref{eq:QrUrprofE_beta,eq:QrUrprofT_beta} as a function of the multipole $\ell$ and the angular distance from the peak $\theta$ to determine which angular distance and multipole ranges contribute the most to the signal. 
This profile depends on the window functions and the power-spectrum profile, with an $\ell$ proportionality for the linear bias, and $\ell^3$ for the scale-dependent bias. 
We thus break the equations into four parts, 
\begin{subequations}
\label{eq:QrUrterms}
    \begin{align}
        A(\ell,\theta)&=\ell W_\ell^E W_\ell^P C_\ell^{EE}J_2(\ell\theta)\, ,\label{eq:lmaxA}\\
        B(\ell,\theta)&=\ell^3 W_\ell^E W_\ell^P C_\ell^{EE}J_2(\ell\theta)\, ,\label{eq:lmaxB}\\
        C(\ell,\theta)&=\ell W_\ell^T W_\ell^P C_\ell^{TE}J_2(\ell\theta)\, ,\label{eq:lmaxC}\\
        D(\ell,\theta)&=\ell^3 W_\ell^T W_\ell^P C_\ell^{TE}J_2(\ell\theta),\label{eq:lmaxD}
    \end{align}
\end{subequations}
\noindent where we have ignored the bias parameters and the cosmic birefringence for this exercise, since they are constants for these equations. 
The window functions here are the $N_\mathrm{side}=2048$ pixel window function, with the \ang{;5;} beam, and the extra \ang{;10;} Gaussian smoothing applied to the data. In \cref{fig:birefcontours} we can see that most of the constraining power lies below $\theta=\ang{2.5}$, and similarly in $\ell$ most of the power is below $\ell=1500$, justifying our choice of cutoff. We find that removing multipoles below $\ell=50$ and choosing an $\ell_\mathrm{max}=2000$ gives negligible changes to the final results.  

\bibliographystyle{JHEP}
\bibliography{birefbib,Planck_bib}

\end{document}